\documentclass[12pt,preprint]{aastex}

\def\msun{$M_{\odot}$}

\def\mdot{$\dot M$}

\def\ergsec{\hbox{erg s$^{-1}$ }}
\def\ergseccm{\hbox{erg s$^{-1} $cm$^{2}$ }}
\def\ergcm{\hbox{erg cm$^{-2}$ s$^{-1}$ }}

\def\Tin{$T_{\rm in}$}
\def\Teff{$T_{\rm eff}$}

\begin{document}

\title{The Spin of the Near-Extreme Kerr Black Hole GRS~1915+105}

\author
{Jeffrey E. McClintock\altaffilmark{1},
Rebecca Shafee\altaffilmark{2},
Ramesh Narayan\altaffilmark{1},
Ronald A. Remillard\altaffilmark{3}, \newline
Shane W. Davis\altaffilmark{4},
Li-Xin Li\altaffilmark{5}}

\altaffiltext{1}{Harvard-Smithsonian Center for Astrophysics, 60 Garden
Street, Cambridge, MA 02138}

\altaffiltext{2}{Harvard University, Department of Physics,
17 Oxford  Street, Cambridge, MA 02138}

\altaffiltext{3}{Kavli Center for Astrophysics and Space Research,
Massachusetts Institute of Technology, Cambridge, MA 02139}

\altaffiltext{4}{Department of Physics, University of California, Santa
           Barbara, CA 93106}

\altaffiltext{5}{Max-Planck-Institut f\"ur Astrophysik,
           Karl-Schwarzschild-Str.\ 1, Postfach 1317, 85741 Garching,
           Germany}

\begin{abstract}

Based on a spectral analysis of the X-ray continuum that employs a fully
relativistic accretion-disk model, we conclude that the compact primary
of the binary X-ray source GRS~1915+105 is a rapidly-rotating Kerr black
hole.  We find a lower limit on the dimensionless spin parameter of $a_*
> 0.98$.  Our result is robust in the sense that it is independent of
the details of the data analysis and insensitive to the uncertainties in
the mass and distance of the black hole.  Furthermore, our
accretion-disk model includes an advanced treatment of spectral
hardening.  Our data selection relies on a rigorous and quantitative
definition of the thermal state of black hole binaries, which we used to
screen all of the available {\it RXTE} and {\it ASCA} data for the
thermal state of GRS~1915+105.  In addition, we focus on those data for
which the accretion disk luminosity is less than 30\% of the Eddington
luminosity.  We argue that these low-luminosity data are most
appropriate for the thin $\alpha$-disk model that we employ.  We assume
that there is zero torque at the inner edge of the disk, as is likely
when the disk is thin, although we show that the presence of a
significant torque does not affect our results.  Our model and the model
of the relativistic jets observed for this source constrain the distance
and black hole mass and could thus be tested by determining a VLBA
parallax distance and improving the measurement of the mass function.
Finally, we comment on the significance of our results for
relativistic-jet and core-collapse models, and for the detection of
gravitational waves.

\end{abstract}



\section{Introduction}

GRS 1915+105 has unique and striking properties that sharply distinguish
it from the 40 known binaries that are believed to contain a
stellar-mass black hole (Remillard \& McClintock 2006, hereafter RM06).
It is the most reliable source of highly relativistic radio jets in the
Galaxy (Mirabel \& Rodr\'iguez 1994; Fender et al.\ 1999; Miller-Jones
et al.\ 2006), and it is the prototype of the microquasars (Mirabel \&
Rodr\'iguez 1999).  GRS~1915+105 (hereafter GRS1915) frequently displays
extraordinary X-ray variability that is not mimicked by any other black
hole system (e.g., Belloni et al.\ 2000; Klein-Wolt et al.\ 2002).  Its
black hole (BH) primary is unique in displaying a constellation of
high-frequency QPOs (HFQPOs), namely, 41~Hz, 67~Hz, 113~Hz and 166~Hz.
The 67~Hz QPO is atypically coherent ($Q \equiv \nu/\Delta\nu \sim 20$)
and relatively strong (rms $ > 1$\%) compared to the HFQPOs observed for
six other accreting BHs (Morgan et al.\ 1997; McClintock \& Remillard
2006, hereafter MR06).  Among the 17 transient and ephemeral systems
that contain a dynamically confirmed BH (RM06), GRS1915 is unique in
having remained active for more than a decade since its discovery during
outburst in 1992 (MR06).  GRS1915 has an orbital period of 33.5 days and
is the widest of the BH binaries (BHBs), and it likely contains the most
massive stellar BH (Greiner et al.\ 2001; Harlaftis \& Greiner 2004;
RM06).

Zhang et al.\ (1997) first argued that the relativistic jets and
extraordinary X-ray behavior of GRS1915 are due to the high spin of its
BH primary.  In their approximate analysis, they found that both GRS1915
and GRO~J1655--40 had high spins, $a_* > 0.9$ ($a_* = cJ/GM^2$, where
$M$ and $J$ are the mass and angular momentum of the BH; $a_* = 0$ for a
Schwarzschild hole and $a_* = 1$ for an extreme Kerr hole).
Subsequently, Gierlin\'ski et al.\ (2001) estimated the spin of
GRO~J1655--40 and LMC X--3.  Recently, we have firmly established the
methodology pioneered by Zhang et al. and Gierlin\'ski et al. by
constructing relativistic accretion disk models (Li et al.\ 2005; Davis
et al.\ 2005) and by modeling in detail the effects of spectral
hardening (Davis et al.\ 2005, 2006).  We have made these analysis tools
publicly available via XSPEC ({\it kerrbb} and {\it bhspec}; Arnaud
1996).  Using this modern methodology, spins have now been estimated for
several stellar-mass BHs, most notably: GRO~J1655--40 and 4U~1543--47
(Shafee et al.\ 2006, hereafter S06), GRS1915 (Middleton et al.\ 2006),
and LMC X-3 (Davis et al.\ 2006).

All of the plausibly reliable estimates of BH spin to date, including
the present work, depend on fits to the X-ray continuum and measurements
of the X-ray luminosity, coupled with optical measurements of BH mass,
orbital inclination, and distance (e.g., S06).  In this paper, we show
that GRS1915 does indeed harbor a rapidly-spinning Kerr BH as suggested
by Zhang et al.\ (1997).  However, in the case of GRO~J1655--40 the
results obtained by ourselves and others show that the spin of this BH
is modest ($a_* \sim 0.75$; S06; Gierlin\'ski et al.\ 2001) and much
lower than the value ($a_* \sim 0.93$) suggested by Zhang et al.  The
high spin reported herein for GRS1915 contradicts the modest spin value
($a_* \sim 0.7$) reported by Middleton et al.\ (2006), and we discuss
this inconsistency in detail in \S5.3.

Our spin estimates are based on an analysis of the ``thermal state'' of
BHBs (MR06) whose remarkably simple properties have been recognized for
decades.  Basic principles of physics predict that accreting BHs should
radiate thermal emission from the inner accretion disk, and a
multi-temperature model of a thin accretion disk was published shortly
after the launch of {\it Uhuru} (Pringle \& Rees 1972; Shakura \&
Sunyaev 1973; Novikov \& Thorne 1973; Lynden-Bell \& Pringle 1974).  A
nonrelativistic approximation to this model, now referred to as {\it
diskbb} in XSPEC (Arnaud 1996) was first implemented and used
extensively by Mitsuda et al.\ (1984) and Makishima et al.\ (1986).  The
two parameters of the model are the temperature $T_{\rm in}$ and radius
$R_{\rm in}$ of the inner edge of the accretion disk.  In their review
on BHBs, Tanaka and Lewin (1995) show for a few BHBs (see their Fig.\
3.14) that as the thermal disk flux varies by 1--2 orders of magnitude
the value of $R_{\rm in}$ remains constant to within $\lesssim 20$\%.
This striking result prompted Tanaka \& Lewin to comment that $R_{\rm
in}$, which was typically found to be $\sim$ tens of kilometers, must be
related to the radius of the innermost stable circular orbit ($R_{\rm
ISCO})$.  The stability of $R_{\rm in}$ has by now been observed in
great detail for many BHBs (e.g., Ebisawa et al.\ 1994; Sobczak et al.\
1999; Sobczak et al.\ 2000; Park et al.\ 2004).  Further strong evidence
for a thermal disk interpretation is provided by plots of the observed
disk flux versus apparent temperature, which track the expected $L
\propto T^4$ relation for a constant inner disk radius (Gierlin\'ski \&
Done 2004; Kubota \& Done 2004).

Spin can be determined because it has a profound impact on the behavior
and properties of a BH.  Quantitatively and specifically, consider two
BHs with the same mass $M$, one a Schwarzschild hole and the other an
extreme Kerr hole.  For the Kerr hole, the radius of the ISCO is six
times smaller and the binding energy at the ISCO seven times greater
than for the Schwarzschild hole.  Relative to the spinless BH, the much
deeper gravity well of the extreme Kerr hole hardens the X-ray spectrum
and greatly increases its efficiency for converting accreted rest mass
into radiant energy.  The continuum fitting approach that we use is
based on measuring spectral shape (hardness) and luminosity
(efficiency).

This paper is organized as follows.  In \S\S2--4 we discuss respectively
the selection, reduction, and analysis of the data.  In \S5 we present
our results for GRS1915 and compare them with those of Middleton et al.\
(2006), and we present a table summary of the spins of GRS1915 and three
other BHs.  The discussion topics in \S6 include a description of our
methodology, our rationale for favoring low-luminosity data, the natal
origin of BH spin, the significance of measuring BH spin, and a proposed
test of our model.  In \S7 we offer our conclusions.


\section{Data Selection}

Our primary resource is the huge and growing archive of data on GRS1915
that has been obtained during the past decade using the large-area PCA
detector on board the {\it Rossi X-ray Timing Explorer} ({\it RXTE};
Swank 1998).  Many BHBs have by now been observed hundreds of times, but
none has been observed more often than GRS1915.  The {\it net} inventory
of {\it RXTE} pointed observations on this source from 1996 to the
present now totals 4.7 Ms, which corresponds to 1311 pointed
observations each of duration 1--10 ks.

The unique properties of GRS1915 and the great volume of perplexing data
present a serious challenge: Is it possible to identify extended periods
of time when GRS1915 was in a genuine pacific state dominated by thermal
emission, and can one use these data to obtain a reliable estimate of
spin?  We believe we have answered ``yes'' to this challenge by using a
quantitative definition of the ``thermal state'' that is based on our
exhaustive studies of many BHBs and BH candidates (MR06; RM06).  For a
discussion of BH states, see MR06 and RM06, and for precise definitions
of the three outburst states -- including the thermal state -- see Table
2 in RM06.  For complete overviews on the evolution and energetics of BH
states for six canonical BHBs (i.e., excluding GRS1915), see \S5 in
RM06.

In the thermal state (formerly high/soft state and ``thermal dominant''
state; MR06), which is the only state relevant to this work, the flux is
dominated by blackbody-like emission from the inner accretion disk, QPOs
are absent or very weak, and the rms variability is also weak.
Quantitatively, the thermal state is defined by two timing criteria and
one spectral criterion applied over the energy band 2--20 keV (MR06;
RM06): (1) QPOs are absent or very weak: amplitude $< 0.005$\%; (2) the
power continuum level integrated over 0.1--10~Hz is $< 0.075$ rms; and
(3) the fraction of the emission contributed by the accretion disk
component $f_{\rm D}$ exceeds 75\% of the total emission.

We now turn to describing how we screened the {\it RXTE} data archive
for GRS1915 and identified 20 observations as belonging strictly to the
thermal state.  As a starting point, all of these individual PCA
observations of GRS1915 that were publicly available as of 2005 January
1 were organized into 640 data segments, where we sometimes combined
brief observations that occurred within an interval of several hours.
We then screened for temporal variability, and 338 relatively ``steady''
observations were identified for which the rms fluctuations in the count
rate divided by the mean count rate was $< 16$\% using 1-s time bins.
Next, a hardness ratio ($HR = 8.6-18.0$ keV / $5.0-8.6$ keV) was
computed for each of these 338 observations using the scheme of Muno et
al.\ (2001) to normalize the PCA count rates for several epochs with
different PCA gain settings.  We then selected a gross sample of 85
observations that displayed the softest spectra ($HR <$ 0.30).  At this
point, we strictly applied the three criteria, which define the thermal
state.  Applying the timing criteria (1) and (2) stated above (i.e., QPO
amplitude $< 0.005$\% and rms continuum power $< 0.075$ rms), left us
with 47 candidate observations.  Finally, based on a decomposition of
the spectrum into thermal and nonthermal components, which is described
in the following section, we obtained our sample of 20 observations that
additionally meets criterion (3) given above, namely, that the thermal
disk component contributes $f_{\rm D} > 75$\% of the total 2--20 keV
flux.  It is this final sample of 20 strictly thermal-state observations
that is the focus of this work.  A catalog of the 20 {\it RXTE}
observations, which span a time interval of 7.5 years, is given in
Table~1.

Finally, we screened the 11 archival observations of GRS1915 obtained by
the {\it Advanced Satellite for Cosmology and Astrophysics (ASCA)} and
identified two appropriate thermal-state observations.  These two
observations, which were made on 1994 September 27 and 1999 April 15,
are also cataloged in Table~1.  In selecting these data we only applied
the spectral criterion (number 3) mentioned above and applied it only
over the observed bandpass of 1.2-10 keV.  The limited count rates
(Table~1) did not allow us to exercise the two timing criteria.  Because
of these limitations, we are somewhat less certain that these
observations correspond to the true thermal state than is the case for
the {\it RXTE} observations.


\section{Data Reduction}

In our spectral analysis of the {\it RXTE} data, we only include
pulse-height spectra from PCU-2 because it is almost always operating
and because fits to the simple power-law (PL) spectrum of the Crab show
that this is the best calibrated proportional counter unit (PCU).  Data
reduction tools from HEASOFT version 5.2 were used to screen the event
files and spectra.  Data were taken in the ``Standard 2 mode,'' which
provides coverage of the PCA bandpass every 16~s.  Data from all Xe gas
layers of PCU-2 were added to make the spectra.  Background spectra were
obtained using the tool {\it pcabackest} and the latest ``bright
source'' background model.  Background spectra were subtracted from the
total spectra using the tool {\it mathpha}.  Redistribution matrix files
and ancillary response files were freshly generated individually for
each PCU layer and combined into a single response file using the tool
{\it pcarsp}.  In fitting each of the 20 pulse-height spectra (\S4), we
used response files that were targeted to the time of each GRS1915
observation.

It is well known that fits to PCA spectra of the Crab Nebula reveal
residuals as large as 1\%, and we therefore added the customary
systematic error of 1\% to all PCU energy channels using the tool {\it
grppha} (e.g., Sobczak et al.\ 2000).  Because large fit-residuals are
often found below 3 keV, which cannot be accounted for by any plausible
spectral feature, and because the spectrum becomes background-dominated
and the calibration less certain above 25 keV, we restricted our
spectral analysis to the 3--25 keV band, which is customary for analysis
of PCA spectra obtained after the gain change of 1999 March.  We used
this same 3--25 keV band even for the 14 PCA data sets that were
obtained prior to 1999 March.

\begin{center}
\begin{tabular}{cccccc}

\multicolumn{6}{c}{Table 1.  Observations of GRS~1915+105} \\ \\


\multicolumn{1}{c}{Mission}&\multicolumn{1}{c}{Obs.}&\multicolumn{1}{c}{Date
(UT)$^a$}&

\multicolumn{1}{c}{MJD}&\multicolumn{1}{c}{Exposure}&

\multicolumn{1}{c}{Count Rate$^b$} \\

\multicolumn{1}{c}{(Detector)}&\multicolumn{1}{c}{No.}&\multicolumn{1}{c}{(yymmdd)}&

\multicolumn{1}{c}{}&\multicolumn{1}{c}{(s)}&

\multicolumn{1}{c}{(counts s$^{-1}$)} \\

\hline
\hline

ASCA & 1 &940927 & 48988.1 & 6019 & 98.3\\
(GIS2) & 2 &990415 & 51283.9 & 7153 & 143.3 \\

\hline
 
RXTE & 1 &960605 & 50239.5 & 10768 & 2197.5 \\
(PCU2) & 2 &960607 & 50241.4 & 10960 & 2356.1 \\
  & 3$^c$ &960703 & 50267.4 & 3424 & 1402.1 \\
  & 4$^c$ &960703 & 50267.5 & 2944 & 1367.0 \\
  & 5 &970819 & 50679.2 & 2176 & 4953.9 \\
  & 6 &970819 & 50679.3 & 2608 & 4686.1 \\
 & 7 &970819 & 50679.4 & 3328 & 5117.6 \\
 & 8 &970819 & 50679.5 & 1488 & 4941.5 \\
 & 9 &971111 & 50763.2 & 10432 & 4532.1 \\
  & 10 &971209 & 50791.2 & 4544 & 5181.7 \\
  & 11 &971211 & 50793.4 & 2368 & 4284.8 \\
 & 12 &980220 & 50804.9 & 5472 & 5426.6 \\
 & 13 &980220 & 50864.9 & 1520 & 2726.2 \\
 & 14$^c$ &980329 & 50901.7 & 2768 & 1282.3 \\
 & 15 &991014 & 51465.6 & 5824 & 4976.6 \\
 & 16 &011024 & 52206.6 & 4384 & 4285.4 \\
 & 17$^c$ &030101 & 52640.4 & 3184 & 1704.0 \\
 & 18 &031029 & 52941.6 & 4128 & 4445.6 \\
 & 19 &031103 & 52946.6 & 2496 & 4594.9 \\
 & 20$^c$ &031124 & 52967.5 & 4240 & 1675.1 \\
\hline 
\hline 
\multicolumn{6}{l}{$^a$ Start time of observation.  MJD = JD --
  2,400,000.5.} \\ 
\multicolumn{6}{l}{$^b$ PCA (full bandwidth): counts s$^{-1}$ per PCU;
1 Crab = 2500 cts s$^{-1}$ per PCU.} \\
\multicolumn{6}{l}{$^c$ Key low-luminosity observations (see \S4.2.1
\& \S6.1).} \\
\end{tabular}
\end{center}

All PCA count rates for the 20 pulse-height spectra were corrected for
dead time.  For all normal events (i.e., good events, rejected events
and events in the propane layer) we adopted a dead time of $\tau_{\rm N}
= 8.83~\mu$s, and for Very Long Events we adopted $\tau_{\rm VLE1} =
59~\mu$s for setting~=~1 and $\tau_{\rm VLE1} = 138~\mu$s for
setting~=~2.  The true event rate corrected for dead time divided by the
observed rate is then $\equiv R_{\rm corr}/R = 1.0 - (R_{\rm N} \times
\tau_{\rm N} + {R_{\rm VLE}}_i \times {\tau_{\rm VLE}}_i$), where the
index $i$ refers to the VLE setting for a given observation.  The dead
time corrections ranged from 1.016 to 1.080.

As in S06, we again found it necessary to correct the effective area of
the PCA despite a recent official correction (Jahoda et al.\ 2006),
which was made using a nominal and approximate spectrum of the Crab
nebula (Zombeck et al.\ 1990).  We have chosen to correct our 3--25 keV
fluxes to the most definitive Crab spectrum available, namely, the PL
index ($\Gamma = 2.10 \pm 0.03$) and normalization ($A = 9.7 \pm 1.0$
ph~cm$^{-2}$~s$^{-1}$) given by Toor \& Seward (1974) and the hydrogen
column given by Willingale et al.\ (2001), which implies a 3.0--25.0 keV
flux of $2.64 \times 10^{-8}$ \ergseccm.  We consider the old Toor and
Seward results more reliable than the current but preliminary results
that are summarized in Kirsch et al.\ (2005).

We made these corrections to the effective area as follows: We selected
25 Crab observations distributed over the 7.5 years spanned by the 20
{\it RXTE} observations.  The Crab pulse-height spectra were corrected
for dead time and joined with their response files in the same manner as
described above for the GRS1915 spectra.  The Crab spectra were fitted
over the range 3--25 keV using a simple PL model with the hydrogen
column fixed at $N_{\rm H} = 3.45 \times 10^{21}$ cm$^{-2}$ (Willingale
et al.\ 2001), and the energy flux was computed over this same interval.
The fluxes so computed systematically exceeded the Crab flux quoted
above by the factor $1.091 \pm 0.013$ (rms).  Therefore, the fluxes we
obtained from the analysis of the 20 spectra (\S4) were all corrected
downward by the reciprocal factor 0.917.

For the two {\it ASCA} spectra (Table~1), we analyzed only the data from
the GIS2 detector; the calibration of the GIS3 detector, in particular
its gain correction, is less certain.  We ignored the data from the SIS
detectors because GRS1915 is bright and the pileup effects are
troublesome (Kotani et al.\ 2000), which makes the SIS data less
suitable for fitting the broad continuum spectrum that is of interest
here.  Starting with the unscreened {\it ASCA} data files obtained from
the HEASARC, we followed as closely as possible the data reduction
procedures and criteria mentioned in Kotani et al.\ (2000).  The GIS
events for each detector were summed within a radius of $6'$ centered on
the source position, and the response function of the X-ray telescope
(Serlemitsos et al.\ 1995) was applied.  Background was not subtracted
for this bright source.  A gain correction based on the instrumental
gold M-edge was applied.  A systematic error of 2\% was added to each
energy channel to account for calibration uncertainties, and the
standard dead time corrections were applied.  No correction to the
effective area is required because the GIS effective-area calibrations
were based on the Toor \& Seward (1974) spectrum of the Crab (Makishima
et al.\ 1996).


\section{Data Analysis}

All of the data analysis and model fitting was performed using HEASOFT
version 5.2 and XSPEC version 12.2 (Arnaud 1996) except for the model
{\it bhspec} (see below), which requires XSPEC version 11.3.  We first
consider the most conventional analysis of all 22 data sets (i.e., 20
{\it RXTE} plus two {\it ASCA}) using the simple multi-temperature disk
blackbody model {\it diskbb} and then describe three successive analyses
of these data sets using our relativistic disk model.

In all the {\it RXTE} spectral fits described herein, we fixed the value
of the hydrogen column density at $N_{\rm H} = 4.0 \times
10^{22}$~cm$^{-2}$.  This value is consistent with the values determined
from an analysis of the {\it ASCA} GIS data for GRS1915 by Ebisawa et
al.\ (1998), who found that $N_{\rm H}$ was ``always within the range
$3.5-4.1 \times 10^{22}$~cm$^{-2}$,'' and by ourselves for observations
\#1 and \#2, respectively: $N_{\rm H} = (3.30 \pm 0.04) \times
10^{22}$~cm$^{-2}$ and $N_{\rm H} = (3.75 \pm 0.04) \times
10^{22}$~cm$^{-2}$ (\S4.1).  Our adopted value of $N_{\rm H}$ is also in
reasonable agreement with the {\it BeppoSAX} value determined by Feroci
et al.\ (1999), $N_{\rm H} \sim 5.6 \times 10^{22}$~cm$^{-2}$, and with
radio and millimeter determinations of the interstellar column, $N_{\rm
H} = (3.5 \pm 0.3) \times 10^{22}$~cm$^{-2}$ (Chapuis \& Corbel 2004).

In the following subsections, we discuss in detail the analysis of the
{\it RXTE} data over the energy range 3--25~keV.  All of these fits
required a nonthermal ``tail'' component of emission plus two additional
weak line and edge components, which are described below.  On the other
hand, the {\it ASCA} GIS pulse-height spectra, which were analyzed over
the energy range 1.2--8~keV required neither a tail component nor the
edge components.  Apart from these simplifications, the only difference
between the analysis of the {\it ASCA} data and the {\it RXTE} data is
that in the former case we allowed $N_{\rm H}$ to vary freely.  Because
of the restricted bandpass of {\it ASCA} and the limitations associated
with screening these data (\S2), we consider the {\it ASCA} results
somewhat less reliable than the {\it RXTE} results, although in the case
of GRO~J1655--40 we found good agreement between the two, most notably
in the case of one simultaneous observation (S06).

\subsection{Nonrelativistic Disk Blackbody plus Simple Power-law Model}

A basic, conventional model consisting of only three principal
components, namely, a multi-temperature disk blackbody ({\it diskbb}), a
simple PL model ({\it power}), and interstellar absorption ({\it phabs})
with $N_{\rm H}$ fixed at $4.0 \times 10^{22}$ (MR06) consistently gave
unacceptably poor fits to the {\it RXTE} data.  In the usual way, we
added two additional components, a Gaussian line {\it gaussian} and a
broad Fe absorption edge ({\it smedge}; e.g., Ebisawa et al.\ 1994;
Sobczak et al.\ 1999, 2000; Park et al.\ 2004; MR06).  In applying the
line component, we followed closely the results obtained from
high-resolution {\it ASCA} SIS observations of GRS1915.  Specifically,
in a pair of GRS1915 SIS spectra, Kotani et al.\ (2000) found a complex
of several, relatively-narrow absorption features that extend from $\sim
6.4-8.3$ keV; for both spectra, the equivalent width of the total
complex is $EW \approx 0.13$ keV.  Accordingly, given the limited
resolution of the PCA ($\approx 18$\% at 6 keV), we added to our basic
model a broad absorption line with a fixed width of 0.5 keV, which we
bounded to lie between 6.3 keV and 7.5 keV.  Then, by adding an
additional broad Fe absorption component (smedge) with an edge energy
restricted to the range 6.9--9.0 keV, we were able to obtain good fits
to all 20 {\it RXTE} spectra.  We note that Kotani et al.\ also used a
sharp absorption edge component in their model, and we used such a
feature in some cases (see \S4.2.4).

Using the model described above, we obtained the values of the
parameters and fluxes plotted in Figure~1.  There are a total of 8 fit
parameters: The disk blackbody temperature \Tin~and its normalization
constant $K$, the PL index $\Gamma$ and its normalization constant, the
smedge optical depth $\tau_{\rm S}$ and the smedge edge energy $E_{\rm
S}$, the central energy of the Gaussian absorption line $E_{\rm Fe}$ and
the intensity of the line $N_{\rm Fe}$.  All the fit parameters, except
for the PL normalization parameter, are shown in Figure~1.  Also shown
is the equivalent width $EW$ of the Gaussian line, the 2--20~keV disk
and PL fluxes ($F_{\rm D}$ and $F_{\rm PL}$, respectively), and the
ratio of these fluxes $f_{\rm D}$, which is a key quantity used in the
selection of these thermal-state data (\S2).

Finally, we briefly summarize our {\it ASCA} GIS2 fit results.  For
observation \#1 (Table 1), we find $kT_{\rm in} = 1.66 \pm 0.03$ keV, $K
= 126.1 \pm 9.3$ and $\chi_{\nu}^{2} = 1.08$ for 98 dof.  For
observation \#2 we find $kT_{\rm in} = 1.91 \pm 0.03$ keV, $K = 137.8
\pm 8.1$ and $\chi_{\nu}^{2} = 0.94$ for 159 dof.  The values of $N_{\rm
H}$ for both observations are quoted above.  A Gaussian absorption line
with a central energy of $6.85 \pm 0.04$~keV and an equivalent width of
0.11~keV was included in the fit to observation \#1, but was not
required or included for observation \#2.  Neither a smedge component
nor a PL or other tail component of emission was included in these
1.2-8.0 keV fits.

\subsection{Relativistic Analysis}

As in S06, we estimate $a_*$ by fitting the thermal component of the
X-ray continuum using a fully relativistic model of a thin accretion
disk around a Kerr BH (Li et al.\ 2005).  The model, which is available
in XSPEC under the name {\it kerrbb}, includes all relativistic effects,
such as frame dragging, Doppler boosting, gravitational redshift, and
light bending.  It also includes self-irradiation of the disk
(``returning radiation'') and the effects of limb darkening.  A
limitation of {\it kerrbb} is that one of its three key fit parameters,
namely, the spectral hardening factor $f$ that relates the color
temperature $T$ and the effective temperature \Teff~of the disk emission
($f$~=~T/\Teff~; Shimura \& Takahara 1995; Merloni et al.\ 2000) is
treated as a constant.

Because of this limitation of {\it kerrbb} our work is based on a
second, complementary relativistic disk model called {\it bhspec}, which
has also been implemented in XSPEC (Davis et al.\ 2005, hereafter D05;
Davis et al.\ 2006, hereafter D06).  It does not include the effects of
returning radiation, but it provides state-of-the-art capability for
computing the spectral hardening factor $f$.  The code {\it bhspec} is
based on non-LTE atmosphere models within an $\alpha$-viscosity
prescription (D05; Shakura \& Sunyaev 1973), has just two principal fit
parameters (spin and mass accretion rate), and can be used directly to
fit for $a_*$ (D06).  As we now describe, our approach is to combine the
functionalities of {\it bhspec} and {\it kerrb} into a single code that
we call {\it kerrbb2}.

The use of this hybrid code {\it kerrbb2} marks an important difference
in methodology between our earlier work (S06) and the present one.  As
discussed in S06, {\it kerrbb} has three fit parameters --- $a_*$, $f$
and the mass accretion rate \mdot~--- only two of which can be
determined at one time.  In S06, we fitted for $f$ and \mdot~with $a_*$
fixed, and we also computed the Eddington-scaled luminosity, $l \equiv
L/L_{\rm Edd}$ [$L_{\rm Edd} = 1.3 \times 10^{38}M$ \ergsec~and $L =
L(a_*,\dot M)$, e.g., Shapiro \& Teukolsky 1984] .  We then plotted $f$
versus $l$ and graphically compared the fit results to a model
calculation of $f$ versus $l$ performed using {\it bhspec}.  Finally, by
varying the assumed value of $a_*$, we determined our estimate of the
spin parameter.  In the present work, this procedure has been
streamlined using {\it kerrbb2}, which we now describe.

The code {\it kerrbb2} is a modified version of {\it kerrbb} that
contains a pair of look-up tables for $f$ corresponding to two values of
the viscosity parameter: $\alpha = 0.01$, 0.1.  The entries in the
tables were computed using {\it bhspec}.  The two tables give $f$ versus
$l$ for a wide range of the spin parameter, $ 0 < a_* < 0.9999$.  The
computations of $f$ versus~$l$ were done using the appropriate,
corresponding response matrices and energy ranges used in fitting the
spectra with {\it kerrbb}.  Thus, {\it kerrbb} and the subroutine/table
computed using {\it bhspec} now allow us to directly fit for $a_*$ and
$l \equiv L/L_{\rm Edd}$ while retaining the special features of {\it
kerrbb} (e.g., returning radiation).  This hybrid code {\it kerrbb2} is
used exclusively in all of the data analysis described herein.

In order to estimate the BH spin by fitting the broadband X-ray
spectrum, one must input known values of the mass $M$ of the BH, the
distance $D$ to the binary, and the inclination $i$ of the black-hole
spin axis, which for GRS1915 we take to be the inclination of the
non-precessing and stable jets (Fender et al.\ 1999; Dhawan et al.\
2000b).  For GRS1915, we adopt the following values for these three
parameters: $M = 14.0 \pm 4.4$~\msun~(Harlaftis \& Greiner 2004), $D =
11.0$~kpc and $i = 66^{\circ} \pm 2^{\circ}$ with $D < 11.2 \pm 0.8$ kpc
(Fender et al.\ 1999).  In this section we use the nominal values of
these parameters, and in \S5.2 we examine the effects on $a_*$ of
allowing these parameters to vary.

In all of the relativistic model fits described below, we used precisely
the same ancillary components with the same constraints that we used in
our nonrelativistic analysis (\S4.1), namely, the 0.5 keV-wide Gaussian
absorption line and the broad absorption component (smedge).
Furthermore, for all of the results presented below, we switched on limb
darkening (lflag = 1) and returning radiation effects (rflag = 1).  We
set the torque at the inner boundary of the accretion disk to zero,
fixed the normalization to 1 (as appropriate when $M$, $i$, and $D$ are
held fixed), allowed the mass accretion rate to vary freely, and fitted
directly for the spin parameter $a_*$.  In the following subsections, we
describe our analysis of the 20 {\it RXTE} and two {\it ASCA} spectra
using {\it kerrbb2} in which we applied in turn three different models
for the tail component, namely, a simple PL model, a thermal
Comptonization model, and a simple PL model plus an exponential cutoff
at lower energies.

\subsubsection{Relativistic Disk plus Simple Power-law Model}

We now consider our baseline analysis of the 20 {\it RXTE} PCA
pulse-height spectra using our relativistic disk model {\it kerrbb2} in
conjunction with a simple power-law component {\it power}.  Following
precisely the prescription we used in our nonrelativistic analysis
(\S4.1), we added two additional components, a broad Fe absorption line
with a fixed width of 0.5 keV (Kotani et al.\ 2000) and a broad Fe
absorption edge (e.g., Ebisawa et al.\ 1994).  These two conventional
and incidental features, which are required in order to obtain a good
fit, are subject to exactly the same constraints as before (\S4.1).  As
stated earlier, these fits were done over the energy range 3--25 keV,
and the column density was fixed to $N_{\rm H} = 4.0 \times 10^{22}$.

As before (\S4.1), there are a total of 8 fit parameters, 6 of which are
identical to those described previously: the PL index $\Gamma$ and its
normalization constant, the smedge optical depth $\tau_{\rm S}$ and the
smedge edge energy $E_{\rm S}$, and the central energy of the Gaussian
absorption line $E_{\rm Fe}$ and the intensity of the line $N_{\rm Fe}$.
Of course, the two principal fit parameters are now $a_*$ and \mdot~in
place of the temperature and disk normalization constant, which are
returned by {\it diskbb}.  The analysis was done for all 20 {\it RXTE}
observations for both values of the viscosity parameter.

The fit results are summarized in Figure 2 ($\alpha = 0.01$) and
Figure~3 ($\alpha = 0.1$) in precisely the same format used in
displaying the {\it diskbb} results in Figure~1.  That is, the structure
of these figures (e.g., the order of parameters and the ranges over
which the parameters are displayed) is identical to the structure of
Figure~1, which summarizes the results of our nonrelativistic analysis
(\S4.1).  There are two important differences to note between Figures 2
\& 3 and Figure~1. First, the obvious difference is that $a_*$ and
\mdot~are now displayed in place of $T_{\rm in}$ and $K$.  Secondly, in
Figures 2 and 3, the value of the disk fraction $f_{\rm D}$ in the top
panel is in the range $f_{\rm D} \sim 0.9-1.0$.  This is generally
significantly greater than the corresponding values of $f_{\rm D}$ shown
in Figure~1, which occasionally dip down to $f_{\rm D} \approx 0.75$.
Thus {\it kerrbb2} is able to accommodate a larger fraction of the total
flux than {\it diskbb} or, correspondingly, the model for the tail
component is less important when fitting with {\it kerrbb2}.

The data points for five of the observations in Figures 2 and 3 are
enclosed by blue circles.  These are the five lowest-luminosity
observations ($L/L_{\rm Edd} < 0.3$).  They are critically important for
our determination of the spin of GRS1915, as we explain in \S6.1 and the
Appendix.  For four of these observations the values of chi-square are
relatively high.  As we show in \S4.2.4, the addition of a minor feature
to the spectral model allows us to obtain good fits ($\chi_{\nu}^{2}
\approx 1$) to these four crucial spectra without significantly
affecting the values of the two important parameters, $a_*$ and \mdot.

Finally, we briefly summarize our {\it ASCA} GIS2 results for the case
$\alpha = 0.01$.  For observation \#1 (Table 1), we find $a_* = 0.988
\pm 0.003$, \mdot~$= (1.40 \pm 0.08) \times 10^{18}$ g~s$^{-1}$, $N_{\rm
H} = (3.39 \pm 0.04) \times 10^{21}$ cm$^{-2}$ and $\chi_{\nu}^{2} =
1.25$ for 95 dof.  For observation \#2 we find $a_* = 0.957 \pm
0.005$,~\mdot $= (3.66 \pm 0.14) \times 10^{18}$ g~s$^{-1}$, $N_{\rm H}
= (3.99 \pm 0.04) \times 10^{21}$ cm$^{-2}$ and $\chi_{\nu}^{2} = 0.82$
for 159 dof.  A Gaussian absorption line with a central energy of $6.77
\pm 0.05$~keV and an equivalent width of 0.21~keV was included in the
fit to observation \#1, but was not required or included for observation
\#2.  No PL or other tail component of emission was included in these
1.2-8.0 keV fits.

\subsubsection{Relativistic Disk plus Comptonization Model}

In the analysis of the {\it RXTE} observations described above in \S4.1
and \S4.2.1, we found that the PL component sometimes makes a modest
contribution to the total flux at energies below $\sim 5$~keV.  We
question whether this contribution from the PL is physically reasonable,
since the PL is believed to be produced by Comptonization of the soft
disk photons by a scattering corona.  In order to check if this PL flux
affects our results, we next fitted the tail component of emission using
a more physically-motivated model for which the disk component dominates
more strongly below several keV.  Namely, we used a thermal
Comptonization model ({\it comptt}) in place of the simple PL component
(Titarchuk 1994; Hua \& Titarchuk 1995).  A drawback of {\it comptt} is
its complexity; it has four principal parameters: the temperature of the
soft input photons $T_{\rm 0}$, the coronal plasma temperature $T_{\rm
cor}$, the optical depth of the corona $\tau_{\rm C}$, and a
normalization parameter.

In determining the spin, we considered three fixed values of $T_{\rm 0}$
(\S5.1) that are centered on 2~keV, which is the nominal value of the
disk temperature determined in \S4.1.  As we show in \S5.1, this choice
is completely unimportant.  We also considered two values of the coronal
temperature, $T_{\rm cor} = 30$~keV and $T_{\rm cor} = 50$~keV, and we
found that this choice is also unimportant.  For the purposes of the
discussion at hand, we adopt the values $T_{\rm 0} = 2.0$~keV and
$T_{\rm cor} = 50$~keV.  Thus, we are left with two fit parameters,
$\tau_{\rm C}$ and the normalization constant.  When fitting with no
constraints on $\tau_{\rm C}$, we found that the parameter sometimes ran
away to unphysically low values ($\lesssim 0.01$).  We therefore set a
hard lower bound on the optical depth: $\tau_{\rm C} > 0.4$ (for $T_{\rm
cor} = 50$~keV).  This bound is based on the values of the photon index
determined in \S4.1 ($\Gamma \lesssim 4$) and a simple calculation that
relies on the Zeldovich approximation as described in \S7.5 of Rybicki
\& Lightman (1979).  Finally, we set {\it comptt's} geometry switch to
$-1$, thereby selecting disk geometry and interpolated values of the
$\beta$ parameter.  Our results for the fitting parameters and other
quantities are summarized in Figure~4 for $\alpha = 0.01$ only.  The
structure of this figure is identical with that of Figure~2 except that
$\Gamma$ is replaced by $\tau_{\rm C}$ and the PL flux $F_{\rm PL}$ is
replaced by the 2--20 keV flux in the {\it comptt} component $F_{\rm
C}$.

\subsubsection{Relativistic Disk plus Cutoff Power-law Model}

Modeling the tail component using the thermal Comptonization model is an
effective way to check on the effects of PL flux below $\sim 5$ keV
(\S4.2.2).  However, this model is quite complex.  Therefore, we now
consider a simpler model that allows us to cut off the flux at low
energy in an ad hoc way, namely, a simple PL model (\S4.2.1) that is
cutoff at lower energies by an exponential ({\it expabs*power} in
XSPEC).  This model has three parameters, the two standard PL parameters
(\S4.2.1) plus a cutoff parameter $E_{\rm c}$.  In \S5.1 we consider
three plausible choices for the cutoff energy ($E_{\rm c} = 8,
10~\&~12$~ keV), but for now we consider only the central value, $E_{\rm
c} = 10$~keV.  The fit results for this simple model are summarized in
Figure~5, which is strictly identical in structure to Figure~2.  The
results shown are for $\alpha = 0.01$.

\subsubsection{Introduction of a Sharp Absorption Edge}

Five values of chi-square in Figure~2 (observation nos. 3, 4, 12, 14 \&
17) are relatively high, $\chi_{\nu}^{2} \gtrsim 1.5$, and the fit to
observation no.\ 14 is unacceptably high, $\chi_{\nu}^{2} = 3.9$ (44
dof).  Furthermore, these same observations give similarly high values
of chi-square for the Comptonization model (Fig.\ 4) and the cutoff PL
model (Fig.\ 5) as well. These particular observations are important
because four of them are low-luminosity observations (\S4.2.1, \S6.1,
Appendix).  In an effort to improve the fits for these five
observations, we followed the lead of Kotani et al.\ (2000; \S4.1).
Specifically, we added to our spectral model a sharp edge feature ({\it
  edge} in XSPEC), which we bounded to lie in the range 8--13 keV, and
we then refitted these five PHA spectra.  The results are summarized in
Figure~6, where the new parameters and fluxes are plotted as red open
circles and the small black data points have been copied from Figure~2.
Apart from the new fit results, Figure~6 differs from Figure~2 in that
it includes a pair of additional panels displaying the parameters of the
edge component, $E_{\rm Ed}$ and $\tau_{\rm Ed}$.  Note in Figure~6 that
the optical depth of the edge component is modest, $\tau_{\rm Ed}
\approx 0.2$, and that the addition of this feature significantly
reduces the optical depth of the smedge component.  Figure~6 contains
two important messages.  First, with the addition of the edge component
all of the five fits are now good ($\chi_{\nu}^{2} \approx 1$).
Secondly, the values of $a_*$ and \mdot~are scarcely affected by the
inclusion of the sharp edge (see \S5.5, Fig.\ 6).  Finally, we found
that the sharp edge gave the same improvements in chi-square and the
same degree of stability in the values of $a_*$ and \mdot~as well when
applied to the Comptonization (\S4.2.2) and cutoff PL models (\S4.2.3).

\subsection{Critique of the Different Analysis Approaches}

The disk fraction $f_{\rm D}$, which is the ratio of the 2--20 keV
thermal disk flux to the flux in the tail component (PL, Compton, or
cutoff PL) is an important parameter and it is therefore displayed in
the top panels in Figures~1--5.  Note that the value of $f_{\rm D}$ in
Figure~1 never dips below 0.75 for any of the 20 observations, which is
a principal selection criterion that we used in selecting these data
(\S2) via the nonrelativistic analysis (\S4.1).  The typical value is
$\approx 90$\%, although for two observations $f_{\rm D}$ does fall
below 80\%.  In the case of the relativistic analyses using the PL tail
model, the values of $f_{\rm D}$ are significantly higher with typical
values $\gtrsim 95$\% and with few values below 90\% (Figures 2, 3 and
5).  The {\it comptt} tail model consistently gives the highest values
of $f_{\rm D}$, which approach 100\%.  In \S4.2.2, we expressed some
reservations about the simple PL component's contribution to the total
flux at low energies.  However, as we show in \S5, our results for the
PL model agree well with the results obtained for the other two tail
models.

A careful comparison of Figures~1--5 shows that the Gaussian line
parameters ($E_{\rm Fe}$, $N_{\rm Fe}$), the line's equivalent width
($EW$), and the smedge parameters ($E_{\rm S}$ and $\tau_{\it S}$)
change very little whether the disk is modeled with {\it diskbb} or with
{\it kerrbb2} and whether the model for the tail component is a simple
PL, a Comptonized plasma, or a cutoff PL.  This strongly indicates that
these ancillary parameters, which are required in order to obtain a good
fit, are quite unimportant. Furthermore, the Gaussian and smedge
components are relatively weak: the Gaussian line has an $EW \approx
0.2$ keV, comparable to the $\approx 0.13$~keV value reported by Kotani
et al.\ (2000), and the optical depth of the smedge component is
moderate, $\tau_{\rm S} \sim 2$ (for comparison, see Ebisawa et al.\
1994; Sobczak et al.\ 1999, 2000; Park et al.\ 2004).
 
Finally, if one considers the principal relativistic fit parameters --
$a_*$ and \mdot~-- plotted in Figures~2--5, one sees that the
corresponding values of these parameters from figure to figure are
little affected by the choice of model for the tail component (i.e., PL,
Compton, or cutoff PL) {or by the inclusion of a sharp absorption edge
(\S4.2.4, Fig.\ 6)}.  Thus, we conclude that our results are robust to
the details of the analysis -- that is, they depend weakly on the line
and edge parameters, and they depend weakly as well on the choice of the
model for the tail component of emission.


\section{Results}

In this section, we present our results in the form of plots of the
dimensionless spin parameter $a_*$ versus the dimensionless luminosity
$l \equiv L/L_{\rm Ledd}$.  The Eddington-scaled luminosity $l$ is
computed from the two {\it kerrbb2} fit parameters $a_*$ and \mdot~and
the BH mass $M$ (\S4.2).  In this section we consider in turn the
following topics: (1) Our results for the spin of GRS1915; (2) the
effects of varying $M$, $i$ and $D$; (3) a comparison of our results
with those of Middleton et al.\ (2006); (4) the effects of returning
radiation and torque; (5) a lower limit on the spin parameter of $a_* >
0.98$; and (6) a comparison of this limit with the spins of three other
sources.

An important point should be mentioned at the outset.  The model that we
employ to fit the continuum spectrum of GRS1915 is physically consistent
only if (i) the accretion disk is in an optically thick thermal state,
and (ii) the disk is geometrically thin in the vertical direction.
Through the stringent data selection described earlier we have ensured
the first requirement, but the second criterion requires a further
restriction of the data.  In \S~6.1 we make use of a Newtonian analysis
to estimate the disk thickness, and in the Appendix we describe a fully
relativistic analysis.  Based on these two analyses, we show that the
accretion disk will be {\it thin} at all radii, with a height to radius
ratio less than 0.1, only if the accretion luminosity is less than 30\%
of the Eddington luminosity.  Only five observations with RXTE and one
observation with ASCA satisfy this restriction, and we therefore focus
most of our attention on these particular data sets (though we present
detailed results for all 22 sets).

\subsection{Spin versus Luminosity for GRS~1915+105}

All the results given in this subsection assume the nominal values of
the optically-determined input parameters given in \S4.2: $M =
14.0~M_{\odot}$, $i = 66^{\circ}$, and $D = 11.0$ kpc (see \S4.2).  In
the following, we show the results of fitting for the spin parameter
using three different tail models in turn -- simple PL, thermal
Comptonization and cutoff PL -- in conjunction with our relativistic
disk model {\it kerrbb2}.

Figure~7 summarizes our fit results (\S4.2.1) obtained using our
baseline PL tail model (MR06; RM06).  The spin parameter is shown
plotted versus the Eddington-scaled luminosity $l$.  The results for all
20 {\it RXTE} and 2 {\it ASCA} observations (Table~1) are included in
this figure.  The results are shown for two value of the viscosity
parameter, $\alpha = 0.01$ and $\alpha = 0.1$.  All of the analyses
reported herein were computed for both values of $\alpha$ ; however, for
low luminosities ($l \lesssim 0.3$), which are strongly favored in this
work (see \S6.1 and the Appendix), the spin estimates are quite
insensitive to the value of $\alpha$ (e.g., Fig.\ 7), and we therefore
generally show results for only $\alpha = 0.01$.  Error bars are
included in Figure~7, although they are generally too small to be
apparent.

The principal result of this paper is captured in the set of six
lowest-luminosity data points (5 {\it RXTE} and 1 {\it ASCA}) in
Figure~7, namely that the spin-parameter estimate is very nearly unity
for $l \lesssim 0.3$.  For the group of four data points at intermediate
luminosities, $0.3 \lesssim l \lesssim 0.45$, the estimated value of the
spin parameter is somewhat depressed, especially for $\alpha = 0.1$.  At
high luminosities, $l \gtrsim 0.65$, the spin estimate is severely
depressed and seen to decrease significantly with increasing $l$. As we
discuss in \S6.1 and the Appendix, there are good reasons to focus only
on those data that correspond to $l < 0.3$.  We thus conclude that
GRS1915 has a spin parameter close to the maximal Kerr value of $a_* =
1$.

We now consider the effects of replacing the simple PL model for the
tail component of emission with a thermal Comptonization model, {\it
comptt}.  As explained in \S4.2.2, we considered this model because we
had reservations about the behavior of the simple PL model at low
energies.  Additionally, a Comptonization model is more physically
motivated and offers a point of comparison with other studies of spin
that exclusively use a supplementary Comptonization model (e.g.,
Middleton et al.\ 2006, D06).  The fitted parameters of this model are
displayed in Figure~8.  As discussed in \S4.2.2, in fitting the data
using this component, we fixed the thermal temperature of the soft seed
photons at three different trial temperatures -- $T_{0}$ = 1.5, 2.0 and
2.5 keV -- where the central value was determined from our
nonrelativistic analysis (\S4.1; Fig.\ 1).  Figure~8 shows $a_*$ versus
$l$ for the three values of $T_{0}$ where it is immediately obvious that
the results obtained using {\it comptt} do not depend on the temperature
of the seed photons over the range considered.  Error bars are
suppressed, but in all cases their extent is less than the height of the
plotting symbols.  The results, which are shown for $\alpha = 0.01$, can
be seen to be nearly identical to the results obtained using the PL
component for $\alpha = 0.01$ (Fig.\ 7).  Again, for $l \lesssim 0.3$ we
find that $a_* \approx 1$ and for intermediate luminosities the spin is
slightly depressed ($a_* \approx 0.98$).  As before, the spin drops very
significantly at high luminosities ($l \gtrsim 0.65$).

Next, we consider the results for the cutoff PL model which, like the
thermal Comptonization model, contributes negligibly to the flux at low
energies.  Relative to the Comptonization model, its chief advantage is
its greater simplicity, and its disadvantage is its lack of physical
motivation (\S4.2).  The results are summarized in Figure~9 for $\alpha
= 0.01$ and for the three values of the break energy mentioned in
\S4.2.2.  The error bars, which do not exceed the size of the plotting
symbols, are suppressed.  As shown in Figure~9, the results are
essentially independent of the choice of cutoff energy.  Furthermore,
the results for the cutoff PL model at both low and intermediate
luminosities are nearly identical to the results obtained with the
thermal Comptonization model (Fig.\ 8) and with the simple PL model for
$\alpha = 0.01$ (Fig.\ 7).

Finally, in Figure~10 we show superposed the results obtained using all
three tail models for $\alpha = 0.01$.  This figure clearly demonstrates
the robustness of our principal result, namely, that the very high spin
of GRS1915 does not depend in any significant way on the model used to
resolve the relativistic disk component from the faint, adulterating
non-disk component.  Furthermore, in \S4.2 we have demonstrated that the
minor fitting components, the Gaussian line and the smedge, operate the
same in all the fits and are therefore incidental to the results that we
have obtained for the spin parameter (Figs.\ 7--10).

\subsection{Effects of Varying $M$, $i$ and $D$ on the Spin of
GRS~1915+105}

Under the assumption of an intrinsically symmetric jet ejection, Fender
et al.\ (1999) place an upper limit on the distance to GRS~1915+105 of
$D = 11.2 \pm 0.8$ kpc.  Further, Fender et al. treat as realistic only
distances that are in the range 9--12 kpc, as indicated by the entries
in their Table~2.  We follow their lead.  As shown in their table, the
kinematic jet model associates with each distance a unique value of the
jet inclination (e.g., adopting the MERLIN values, $D = 11$ kpc
corresponds to $i = 66^{\circ}$), which we take as the spin axis of the
BH and the accretion disk (\S4.2).  In turn, each value of $i$ is
associated with a definite value of the BH mass via the dynamical
results for GRS1915 (Greiner et al.\ 2001; Harlaftis \& Greiner 2004).
Thus, we have a correlated triplet of numbers $D$, $i$ and $M$, which
are given in Table~2 for five values of $D$.  In Figure~11$a$, we show
the effects of varying $D$ from 11--12.5 kpc.  The cases $D = 9-10$~kpc
are not shown because these values drive $a_*$ toward higher values and
we are interested here in highlighting the lowest values of $a_*$.
Furthermore, as discussed in \S6.4, our fit results indicate that the
distance to GRS1915 is unlikely to be less than 9--10 kpc.  In
Figure~11$a$ and Table 2, we also include $D = 12.5$~kpc because this
extreme distance was adopted by Middleton et al.\ (2006).

\begin{center}
\begin{tabular}{ccc}
\multicolumn{3}{c}{Table 2.  Parameters for GRS 1915+105} \\ \\
\hline
\hline
\multicolumn{1}{c}{Distance$^{a}$}&\multicolumn{1}{c}{Inclination$^{a}$}&\multicolumn{1}{
c}{Mass$^{b}$} \\
\multicolumn{1}{c}{(kpc)}&\multicolumn{1}{c}{(degrees)}&\multicolumn{1}{c}{($M_{\odot}$)}
 \\
\hline
\hline
9.0  & 61.5 & 15.5 \\
10.0 &63.9 & 14.6\\
11.0 & 66.0 & 14.0 \\
12.0 & 67.8 & 13.5 \\
$12.5^{c,d}$ & $68.6$ & 13.3 \\
\hline
\hline
\multicolumn{3}{l}{$^{a}$ Fender et al.\ 1999.} \\
\multicolumn{3}{l}{$^{b}$  Based on $P_{\rm orb}$, $K_2$ and $M_2$ from
Harliftis \& Greiner 2004.} \\
\multicolumn{3}{l}{$^{c}$ Adopted by Middleton et. al.\ 2006.} \\
\multicolumn{3}{l}{$^{d}$ Intrinsic jet velocity $ > c$.} \\
\end{tabular}
\end{center}

In addition to the uncertainty in the distance, the
dynamically-determined value of the BH mass carries its own sizable
uncertainty, $M = 14.0 \pm 4.4$~\msun, because the radial velocity
amplitude of the secondary is known only to a precision of 11\% (Greiner
et al.\ 2001; Harlaftis \& Greiner 2004).  The effects on the spin due
to this uncertainty in the mass are shown in Figure~11$b$.  As indicated
in the figure, the smallest mass, $M = 9.6$, gives the lowest values of
spin.  Limiting our consideration to $L/L_{\rm Edd} < 0.3$ (\S6.1,
Appendix), Figure~11$ab$ shows that for most allowable distances and
masses the spin parameter is nearly unity (see \S5.5).  Finally, on a
separate and incidental matter, we note that our {\it RXTE} results are
also insensitive to our adopted value of $N_{\rm H}$ over the full range
indicated (\S4) because the absorbing column is already $\sim85$\%
transmitting at the PCA's detector threshold energy of 3 keV.

\subsection{Comparison with the Results of Middleton et al. for
GRS~1915+105}

For the nominal 11 kpc distance that we adopt (\S5.2), Middleton et al.\
(2006) report a single, moderate value of the spin parameter of $a_*
\sim 0.8$ (or $a_* \sim 0.7$ for their adopted distance of 12.5 kpc).
As we make clear in \S5.5, the M06 value of $a_* \sim 0.8$ is very much
less than the value we find: $a_* \sim 0.98-0.99$.  M06 and we used
precisely the same publicly-available relativistic accretion disk models
(i.e., {\it kerrbb} and {\it bhspec}).  The key difference between the
two studies is in the methodology of data selection.  M06 used a quite
different approach that yielded a restricted data sample comprised
solely of high-luminosity observations.  As we conclude below, our
results are in fact in reasonable agreement with Middleton et al. in
this high-luminosity regime, which we argue is unreliable for the
determination of spin (\S6.1, Appendix).

Both M06 and we agree completely on the necessity of selecting spectra
that minimize the nonthermal component and that are dominated by disk
emission.  However, M06's methodologies for selecting such spectra were
quite different from ours (\S2).  One difference is that M06 based their
initial selection on the state classifications defined by Belloni et
al.\ 2000 (see also Belloni et al.\ 1997, and Belloni 2004), which were
devised primarily to study disk-jet coupling via a unified model of
X-ray states and radio jets.  We, on the other hand, used quantitative
state definitions that are centered on physical models of X-ray states
(MR05; MR06).  These latter state definitions have been applied more
widely to many BH binaries including GROJ~1655--40, 4U~1543--47,
XTE~J1550--564, H~1743--322, XTE~J1859+226, and GX~339--4 (MR05; MR06).
A second difference is that we screened all the available data and
identified 22 observations that are strictly thermal-state data.  This
yielded a total of 89 ks of {\it RXTE} data and 13 ks of {\it ASCA} data
(Table~1) compared with the much smaller collection of data considered
by M06.  It is this larger data sample that allowed us to identify
several crucial observations at low luminosities ($L/L_{\rm Edd} < 0.3$)
that are completely absent in M06's data sample.

In the end, M06 fitted jointly three representative 16-s observations in
order to determine a single spin estimate with the nominal value of $a_*
= 0.82$ (for $D = 11$~kpc).  This single spin value and the three
corresponding luminosities are indicated in Figure~12 by the three
filled triangles, which are connected by a solid line.  As shown, these
three luminosities range from $L/L_{\rm Edd} = 0.40 - 1.45$.  Note that
MR06's low value of spin, $a_* \sim 0.82$, is in reasonable accord with
the value of $\approx 0.88$ that we find (Fig.\ 12) for a luminosity of
$\approx 80$\% of $L_{\rm Edd}$ using the {\it comptt} model, which is
analogous to the {\it thcomp} tail model that M06 used.  Furthermore,
their somewhat lower spin value, which is an average over a wide range
of luminosity, may be largely due to the inclusion of an observation at
super-Eddington luminosity ($L/L_{\rm Edd} \approx 1.45$; Fig.\ 12).
Note also that even their mid-luminosity observation with $L/L_{\rm Edd}
\approx 0.85$ (Fig.\ 12) corresponds to the effective Eddington
luminosity for thin-disk geometry (\S6.1, Appendix).  As already
mentioned, and discussed in further detail in \S6.1 and the Appendix,
the continuum spectral models used by M06 and us are not self-consistent
and become progressively less reliable at higher luminosities.  In fact,
all three 16-second observations of M06 correspond to $l > 0.3$ and are
thus in a regime where a number of the physical assumptions which
underly the spectral models are likely to break down.

\subsection{Effects of Returning Radiation and Nonzero Torque at the ISCO}

All of our results include the self-irradiation of the disk as a result
of light deflection (assuming that the disk is infinitely thin, see Li
et al.\ 2005), which we refer to as returning radiation.  The effects on
our results of turning off the returning radiation (rflag = 0) is shown
in Figure 13.  As indicated in the figure, the returning radiation
boosts the luminosity of the disk by several percent, but has no
significant effect on the spin parameter.  The returning radiation
feature is not included in {\it bhspec}, the relativistic disk model we
used to compute tables of the spectral hardening factor, which were
incorporated into {\it kerrbb} via a subroutine to create {\it kerrbb2}
(\S4.2).  Both {\it kerrbb2} and {\it bhspec} can be used directly to
determine the principal fit parameters $a_*$ and \mdot.  We made a
thorough comparison of the fit results obtained using the two models for
GRS1915 (and for 4U~1543--47 and GRO~J1655--40 as well).  For the
purposes of this comparison only, we switched the returning radiation
off for {\it kerrbb2} (rflag = 0).  We found that the two models gave
very comparable results for $a_*$ versus $l$.

Throughout this paper we assume that there is no torque acting at the
inner edge of the disk.  This assumption is in agreement with the
classic and current literature on thin-disk accretion, which advocates
the use of a zero-torque boundary condition (Shakura \& Sunyaev 1973;
Novikov \& Thorne 1973; Afshordi \& Paczy\'nski 2003; Li 2003).
However, as discussed in \S6.1, a torque may be present near the ISCO,
especially in the case of thicker, higher-luminosity disks.  Our model
{\it kerrbb2} is quite general and capable of handling positive torques
of any magnitude with the dimensionless torque parameter $\eta_{\rm T}$
defined as the ratio of the power generated by the torque to the
gravitational binding energy of the accreted gas (Li et al.\ 2005).  As
illustrated in Figure 14, the spin parameter decreases with increasing
torque.  In the presence of sizable torques, the spin parameter of
GRS1915 is significantly depressed at high luminosities, but it is
scarcely affected at low luminosities.

\subsection{Summary of Results for GRS~1915+105 and Three Additional 
Sources}

In Table~3 we summarize the average values of the spin of GRS1915
returned by {\it kerrbb2} for $L/L_{\rm Edd} < 0.3$ that are based on
the nominal values of $M$, $i$ and $D$ (\S4.2) and on $\alpha = 0.01$.
The fit results are given for each of the three tail models (\S4.2).
The quantities displayed are the Gaussian-weighted mean value of the
spin $a_*$ and the standard deviation for $N = 5$.  As indicated by
comparing the two lines in the table, the inclusion of a sharp
absorption edge in the spectral model (\S4.2.4) has a negligible effect
on the value of the spin parameter.

\begin{center}
\begin{tabular}{ccccccc}
\multicolumn{7}{c}{Table 3: Fitted values of spin for $L/L_{\rm Edd}< 0.3$} \\ \\
\hline
\hline
\multicolumn{1}{c|}{Object}&\multicolumn{6}{c}{Model} \\
\hline
\multicolumn{1}{c|}{}&\multicolumn{2}{c|}{Power
  Law}&\multicolumn{2}{c|}{Comptt}&\multicolumn{2}{c}{Cutoff Power
  Law}\\
\hline
\multicolumn{1}{c|}{}&

\multicolumn{1}{c|}{mean}&
\multicolumn{1}{c|}{st. dev.}&

\multicolumn{1}{c|}{mean}&
\multicolumn{1}{c|}{st. dev.}&

\multicolumn{1}{c|}{mean}&
\multicolumn{1}{c}{st. dev.} \\
\hline
\hline
GRS 1915+105$^a$  & 0.998&0.001 & 0.997&0.001 & 0.997&0.001\\
GRS 1915+105$^b$  & 0.998&0.001 & 0.995&0.002 & 0.996&0.001\\
\hline
\hline
\multicolumn{7}{l}{$^a$ Sharp absorption edge excluded from the fit;
  see \S4.2.4.} \\
\multicolumn{7}{l}{$^b$ Sharp absorption edge included in the fit; see
  \S4.2.4.} \\
\end{tabular}
\end{center}

The formal and precise values of $a_*$ in Table~3 for all three models
(with and without the edge) are consistent with the physical limit on
the Kerr parameter of $a_* = 0.998$ computed by Thorne (1974).  We
consider this agreement accidental given the likely uncertainties in the
idealized thin-disk model and the model for spectral hardening, the
systematic uncertainties in the data, and the uncertainties in $M$, $i$
and $D$.  Nevertheless, the results in Table~3 indicate a very high
value for the spin parameter of $a_* \gtrsim 0.99$.

We now consider somewhat lower values of spin that cropped up during our
analysis.  We restrict our discussion to $L/L_{\rm Ledd} < 0.3$ (\S6.1,
Appendix).  For example, for the first {\it ASCA} observation (Table 1),
we find $a_* = 0.988 \pm 0.003$ for nominal values of $M$, $i$ and $D$
(\S4.2.1).  Figure~11$a$ and 11$b$ show respectively the effects of
changing $D$ and $M$ on $a_*$.  Considering distance, the spin is lowest
for $D = 12.5$~kpc: $a_* = 0.991 \pm 0.006$ (weighted mean for three
observations).  Considering mass, the spin is lowest for $M =
9.6$~\msun~(single observation with $a_* = 0.987 \pm 0.001$).  Finally,
the values of $a_*$ are slightly less if one considers the case $\alpha
= 0.1$ (Fig.\ 7).  Based on these and other considerations and the
results in Table~3, we adopt $a_* > 0.98$ as a lower limit on the spin
parameter.

This lower limit of 0.98 for the spin of GRS1915 and our previously
estimated values of $a_*$ for GRO~J1655--40 and 4U~1543--47 (S06) are
summarized in Table~4.  We obtained very similar estimates of $a_*$ for
the latter two sources using our revised code {\it kerrbb2} (\S4.2), and
we will report on this work in a later paper.  Also given in Table 4 is
an upper limit on $a_*$ for LMC~X-3 obtained by D06.  Here we provide
conservative estimates of spin, which are based on the considerations
given above for GRS1915 and the full range of variation of $a_*$
considered for GRO~J1655--40 and 4U~1543--47 in R06.  The spin estimates
given in Table~4 are our bottom-line results.

\begin{center}
\begin{tabular}{ccc}
\multicolumn{3}{c}{Table 4. Spin estimates for four sources} \\ \\
\hline
\hline
\multicolumn{1}{c}{Black
Hole}&\multicolumn{1}{c}{Mission}&\multicolumn{1}{c}{$a_*$}
\\ \hline \hline 
GRS 1915+105 & RXTE/ASCA     & $ > 0.98$  \\ 
GRO J1655--40 & RXTE/ASCA     & 0.65-0.75 \\ 
4U 1543--47   & RXTE          & 0.75-0.85  \\  
LMC X--3      & RXTE/BeppoSAX & $< 0.26^a$ \\ 
\hline
\hline
\multicolumn{3}{l}{$^{a}$ Davis et al.\ 2006.}\\
\end{tabular}
\end{center}

It is important to emphasize that the spins of GRO~J1655--40 and
4U~1543--47, although sizable, are effectively very much less than that
of GRS1915, which in turn is significantly less than the theoretical
maximum value of $a_* = 1$.  The implications of the extreme spin of
GRS1915 are not immediately apparent if one considers the parameter
$a_*$ alone.  Therefore it is instructive to consider such related
dimensionless parameters as the radius of the innermost stable circular
orbit (ISCO) $\xi$, the binding energy per unit mass at the ISCO $\eta$,
and the Keplerian frequency at the ISCO $\omega_{\rm K}$, which are all
monotonic functions of $a_*$ (e.g., Shapiro \& Teukolsky 1984).  These
three quantities are defined and plotted versus $a_*$ in Figure~15,
which also shows for the four BHs in question the values of these
quantities for our nominal estimates of spin.  In this approximate
comparison, note that both the nominal Keplerian frequency and the
binding energy at the ISCO for GRO~J1655--40 and 4U~1543--47 are only
half the values indicated for GRS1915.


\section{Discussion}

There are four avenues for measuring spin -- continuum fitting,
high-frequency QPOs, the Fe K line, and polarimetry (RM06).  Because
spin is such a critical parameter it is important to attempt to measure
it by as many of these methods as possible, as this will provide
arguably the best possible check on our results.  The best current
method, continuum fitting, has the drawback that its application
requires accurate estimates of BH mass $M$, disk inclination $i$, and
distance $D$.  In contrast, observations of HFQPOs require knowledge of
only $M$ to provide a spin estimate, and once the correct model is known
this method is likely to offer the most reliable measurements of spin.
Presently, however, the leading model of HFQPOs, which was initially
proposed by Abramowicz \& Kluz\'niak (2001), does not provide a useful
constraint on $a_*$ for GRS1915 because of the wide range of possible
resonances and the sizable uncertainty in the BH mass (T\"or\"ok et al.\
2005).  Another HFQPO model, on the other hand, predicts the precise
value $a_* = 0.99616$ for the spin of GRS1915 (Aschenbach 2004).
Broadened iron lines do not even require $M$, although knowledge of $i$
is useful in order to avoid having to include that parameter in the fit.
However, there are serious sources of uncertainty in the model,
including the placement of the continuum, the model of the fluorescing
source, and the ionization state of the disk (Reynolds \& Nowak 2003).
Furthermore, in the case of GRS1915, the line is seldom seen and has
provided poor constraints on the models, and no estimate of spin has
been given (Martocchia et al.\ 2002, 2004; Miller et al.\ 2004).
Polarimetry appears promising because the polarization features of BH
disk radiation can be affected strongly by GR effects (Lightman \&
Shapiro 1976; Connors et al.\ 1980; Dov\v{c}iak et al.\ 2004).
Unfortunately, however, there have been no such observations of BHBs,
and there are no mission opportunities on the horizon.  In short, the
HFQPO and Fe-line methods are not well enough developed to provide
dependable results, and the required polarimetry data are not available,
whereas the continuum method, despite its limitations, is already
delivering results.

The methodology of the continuum-fitting approach is straightforward and
transparent.  Its foundation is (1) the simplicity of the rigorously
defined thermal state (\S1 \& \S2), which matches very closely the
predictions of the classic thin disk models (\S1), and (2) the vast
amount of X-ray spectral data contained in the NASA/GSFC HEASARC
archives for missions ranging from {\it Ginga} and {\it RXTE} to {\it
Chandra} and {\it XMM-Newton}.

In the thin disk model, there is an axisymmetric radiatively-efficient
accretion flow in which, for a given BH mass $M$, mass accretion rate
$\dot M$ and BH spin parameter $a_*$, one can calculate very accurately
the total luminosity of the disk, $L_{\rm disk} = \eta\dot Mc^2$.  The
parameter $\eta$, which measures the radiative efficiency of the disk,
is a function only of $a_*$ (see Fig.\ 15).  We can also calculate
precisely the local radiative flux $F_{\rm disk}(R)$ emitted at radius
$R$ by each surface of the disk.  Moreover, the accreting gas is
optically thick, and the emission is thermal and blackbody-like, making
it straightforward to compute the spectrum of the emission.  Most
importantly, the inner edge of a thin disk is located very close to the
innermost stable circular orbit (ISCO) of the BH spacetime, whose radius
$R_{\rm ISCO}$ (in gravitational units) is a function only of the spin
of the BH: $R_{\rm ISCO}/(GM/c^2) = \xi(a_*)$, where $\xi(a_*)$ is a
monotonically decreasing function of $a_*$ (see Fig.\ 15).  Thus, if one
measures the radius of the disk inner edge, and if one also has an
estimate of the mass $M$ of the BH, then one can immediately obtain
$a_*$.  This is the principle behind our method of estimating BH spin,
which was first described by Zhang et al.\ (1997).

There is one principal difficulty in applying this method. At the high
disk temperatures typically found in BHB disks (\Tin~$\sim 10^7$~K), the
spectral hardening factor $f$ (\S4.2) is expected to deviate
substantially from unity.  It is therefore important to have a reliable
estimate of $f$.  Until recently, the only estimate available was that
from Shimura \& Takahara (1995), whose seminal but limited study was
rather approximate.  Within the last year, D05 along with Davis \&
Hubeny (2006) have computed more accurate disk atmosphere models
including metal opacities and have obtained reliable estimates of $f$ as
a function of the disk luminosity and inclination.  The use of a
rigorous and modern estimate of $f$ is absolutely essential for the
successful application of this method of estimating BH spin, and it is
only now that such an estimate has become available.  Nevertheless, even
at the lower luminosities we favor (\S6.1), the vertical structure of
real magnetohydrodynamical (MHD) disks may differ in detail from our
models (see \S3 of D06 for details).  However, preliminary
investigations which incorporate the results of MHD simulations suggest
only small changes ($\Delta f/f \lesssim 15$\%).

\subsection{Rationale for Reliance on Low-Luminosity Data}

In S06, we argued that the method employed in that paper as well as the
present paper to estimate BH spin is most reliable at low disk
luminosities.  The argument has been amplified by D06 (see their \S3.1).
The main reason to distrust high luminosity data is that the disk is
likely to be vertically thick, whereas the model explicitly assumes a
thin disk.  The detailed general relativistic ray tracing used in {\it
kerrbb}, {\it kerrbb2} and {\it bhspec} assumes a razor-thin disk whose
surface is exactly at $z=0$.  So long as the disk thickness $H$ is much
less than the local disk radius $R$, we expect only small errors to
result from the idealized geometry assumed in the model.  However, as
$H/R$ increases we expect various geometrical effects to creep in.
Although it is hard to be quantitative, it is reasonable to think that
the errors will become non-negligible once $H/R > 0.1$.

Another important assumption in the models is that there is no torque
applied at the inner edge of the disk (\S5.4).  Krolik (1999) and Gammie
(1999) argued that magnetic fields would be amplified near the inner
edge of the disk, where the gas begins to free-fall into the BH, and
that these fields would apply a torque on the disk.  The torque will
enhance the energy dissipation near the ISCO and lead to a modification
in the profile of the disk flux $F(R)$.  If this effect is strong enough
it will introduce a large error in the BH spin estimate.  Interestingly,
Li (2004) finds that, in some cases, a strong magnetic field connecting
the disk and the BH may actually move the inner edge of the disk out and
cause a reduction in the luminosity.  Afshordi \& Paczy\'nski (2003)
suggested that the torques are likely to scale as some positive power of
$H/R$ and therefore will be unimportant in very thin disks (see their
Figs.\ 17 and 18).  This topic is still under debate and is likely to be
settled only with detailed models.

To make progress on this question, one approach is to work with the
viscous hydrodynamic disk equations, including pressure and radial
dynamics (as in Narayan, Kato \& Honma 1997 and Afshordi \& Paczy\'nski
2003), and to calculate the viscous stress at the sonic radius and the
rate of viscous energy dissipation as a function of radius for various
disk thicknesses.  Within the limitations of the $\alpha$-viscosity
prescription, this will provide a clean answer to whether or not the
torque at the inner edge is important for thin disks.  Our preliminary
analysis appears to support Afshordi \& Paczy\'nski's (2003) assertion
that the torque is unimportant for thin disks.  A more detailed, and
ultimately more rigorous, approach is to carry out 3D numerical MHD
simulations of realistic thin disks, including radiative cooling to keep
the disk thin.  The only work to date involves non-radiative thick disks
and is not yet very useful.

Regardless of the current uncertainty over the magnitude of the torque
at the disk inner edge, we note that at low luminosities (when the disk
is thin, see below) the effects on the spin parameter of even a sizable
torque is quite small (see \S~5.4).

Another effect that becomes important when the disk is vertically thick
is radial advection of energy (Abramowicz et al.\ 1988, 1995; Narayan \&
Yi 1994, 1995).  The more energy advection there is in the disk, the
less energy is radiated to infinity, and the larger is the deviation of
$F(R)$ from the idealized thin disk profile assumed in the model.  Thus,
it is safest to work with disks that have negligible radial advection,
i.e., disks with $H/R \ll 1$.

Let us now estimate the disk thickness $H$ and the ratio $H/R$ for a
Newtonian thin accretion disk.  The flux emitted by a thin accretion
disk around a BH with zero torque at the inner edge is given by (e.g.,
Frank et al.\ 2002)
\begin{equation}
F(R) = {3GM\dot M\over 8\pi R^3} \left[1-\left({R_{\rm in}\over
R}\right)^{1/2}\right],
\end{equation}
where $M$ is the mass of the BH, $\dot M$ is the mass accretion rate,
$R$ is the cylindrical radius, and $R_{\rm in}$ is the radius of the
inner edge of the disk.  Let us define the Eddington mass accretion
rate by equating the disk luminosity to the Eddington luminosity,
\begin{equation}
{GM\dot M_{\rm Edd}\over 2R_{\rm in}} = L_{\rm Edd} \equiv {4\pi
GMc\over \kappa}, \qquad {\rm i.e.,} \qquad \dot M_{\rm Edd} = {8\pi c
R_{\rm in}\over\kappa},
\end{equation}
where $\kappa$ is the opacity of the gas.  Correspondingly, let us
define the Eddington-scaled mass accretion rate by
\begin{equation}
\dot m \equiv {\dot M\over \dot M_{\rm Edd}}.
\end{equation}
We now rewrite the disk flux $F(R)$ in terms of $\dot m$ and calculate
the vertical acceleration due to radiation pressure,
\begin{equation}
g_{\rm rad}(R) = {F(R)\kappa\over c} = {3GM\dot m\over R^2}
\left[\left({R_{\rm in}\over R}\right) -\left({R_{\rm in}\over
R}\right)^{3/2}\right].
\end{equation}
In equilibrium, the radiative acceleration must be balanced by the
vertical component of gravity, which for simplicity we can write as
$g_z(R,z) = GMz/ R^3$.  We then find
\begin{equation}
{H\over R} \approx 3\dot m
\left[\left({R_{\rm in}\over R}\right) -\left({R_{\rm in}\over
R}\right)^{3/2}\right]. \label{HoverR}
\end{equation}
Figure 16 shows $H/R$ as a function of $R$ for various choices of the
accretion rate: from below, $\dot m = 0.1$, 0.2, ..., 1.2.  If we wish
to have $H/R < 0.1$ at all $R$, then we see that we are limited to
$\dot m \lesssim 0.25$, i.e., to Eddington-scaled disk luminosities $l
\lesssim 0.25$.  The Appendix describes a more accurate estimate of
the disk height that is calculated for a general relativistic disk around a
Kerr BH.  Results are shown in Figure~17.  According to that analysis,
in order to have $H/R \lesssim 0.1$ at all radii, we must restrict our
attention to luminosities $l \lesssim 0.3$.  This is the limit we
employ throughout the paper.

If we consider the exact expression for the Newtonian vertical gravity
\begin{equation}
g_z(R,z) = {GMz \over (R^2+z^2)^{3/2}},
\end{equation}
rather than the approximation $GMz/R^3$, then we find that the maximum
value of the vertical gravitational acceleration (which is achieved at
$z=R/\sqrt{2}$) is
\begin{equation}
(g_z)_{\rm max} = {2\over3\sqrt{3}} {GM\over R^2}.
\end{equation}
For any accretion rate greater than about 85\% of Eddington, or
$\log(L/L_{\rm Edd})>-0.06$, one finds that some parts of the disk
produce too much radiation to be balanced even by the maximum vertical
gravity $(g_z)_{\rm max}$.  Radiation pressure will then cause
material to be blown away from the disk.  This critical luminosity is
clearly related to the Eddington limit; the slightly different
numerical value, i.e., 85\% instead of 100\% of the canonical
Eddington limit, is the result of the different geometry of a disk
compared to the spherical geometry that one usually considers.  (See
Nityananda \& Narayan 1982 for a discussion of geometry effects on the
Eddington limit.)

We showed in \S~5 that, by focusing on data corresponding to $L/L_{\rm
Edd} \lesssim 0.3$, we obtain very consistent results for the spin
parameter of GRS1915, independent of the details of the spectral model
we employ.  We also found that the results begin to deviate as we go to
higher luminosities, suggesting that as the disk thickens one or more of
the effects described in this subsection becomes important.  It is
interesting that the deviations are not random but very systematic,
e.g., the estimate of $a_*$ decreases smoothly and monotonically as
$L/L_{\rm Edd}$ increases.  This signature could conceivably be used to
identify which of our assumptions breaks down as the luminosity
increases.  Detailed viscous disk models with varying disk thickness
might be able to shed some light on this issue.

\subsection{The Spins of Stellar Black Holes are Chiefly Natal}

King and Kolb (1999) provide a global evolutionary and observational
argument that neither significant spinup nor spindown is likely to occur
during the lifetime of any BH binary and hence that BH primaries
essentially retain the spin rates that they had at birth.  In the
particular case of 4U~1543--47, based on its present accretion rate and
modest age ($\lesssim~1$~Gyr), we argued that the spin of its BH (\S5.7)
is likewise chiefly natal (S06).  The fast spin reported herein for
GRS1915, $a_* > 0.98$, is almost certainly a natal spin because the
alternative, achieving this spin gradually via accretion torques, would
require almost doubling the mass of the BH (see below).  Such a large
increment in BH mass is unlikely to have occurred during the evolution
of GRS1915 or any BH binary simply because systems with initially low-
or moderate-mass secondaries (i.e., $M \lesssim$ a few \msun) obviously
cannot supply the required mass, and systems with high-mass secondaries
have lifetimes that are too short to effect the required mass transfer.
We now consider the exceptional case of GRS1915 in more detail.

GRS1915 presently has a low-mass secondary, $M_2 = 0.81 \pm
0.53$~\msun~(Harliftis \& Greiner 2004) and the most massive primary and
longest period of any BH binary (\S1).  There is a great deal of
uncertainty in evolutionary models for GRS1915 and for all BH binaries.
The specific evolutionary model of Belczynski \& Bulik (2002) for
GRS1915 argues for a small transfer of mass to the primary and
negligible spin up, which is in agreement with most generic models
(e.g., King \& Kolb 1999).  An evolutionary model of GRS1915 that links
this source to the ultraluminous X-ray sources implies the most extreme
mass transfer and spin up (Podsiadlowski et al.\ 2003).  These authors
argue that the initial secondary mass could have been as high as 6
\msun~and the BH primary could have accreted as much as $\sim
4$~\msun~(see also Lee et al.\ 2002).  Even for this extreme scenario,
the predicted spin up due to accretion torques is modest. Based on a
precise calculation that ignores returning radiation (\S5.5), we find
that the transfer of 4 \msun~onto a 10 \msun~natal black hole with zero
initial spin yields a final spin of only $\sim 0.77$, which is far less
than our limit of $a_* > 0.98$ (\S5.5, Fig.\ 15).  Likewise, to achieve
a final spin of $a_* > 0.98$ would require an initial spin of $a_* >
0.75$.  Furthermore, if one includes the effects of returning radiation,
then the accretion is less efficient in spinning up the hole and a
somewhat larger natal spin is required.  Again neglecting returning
radiation, a 10 \msun~BH that is spun up by accretion torques from $a_*
= 0$ to $a_* = 0.98$ would have a final mass of 19.3 \msun; since some
of the rest mass energy is radiated away, the total rest mass accreted
in such a spin up event would be 10.7 \msun.  We thus conclude that the
extreme spin of GRS1915 was likely imparted to the BH primary during the
process of its formation.

The generation of large spins is central to GRB models.  Natal spins of
$a_* \sim 0.8$, in agreement with our observations, were predicted for
GRO~J1655--40 and 4U~1543--47 by Lee et al.\ (2002).  The extreme spin
of GRS1915, $a_* > 0.98$, is an expected consequence of collapsar models
(\S6.3).

\subsection{Significance of Measuring Black Hole Spin}

The properties of a BH are completely defined by specifying just two
parameters, its mass $M$ and its dimensionless spin parameter $a_*$.
Furthermore, a BH's mass simply supplies a physical scale, whereas its
spin fundamentally changes the geometry of space-time.  Accordingly, in
order to model the ways in which an accreting BH can interact with its
environment, one must know its spin.  For example, consider one of the
most intriguing unsolved problems in astrophysics, namely, the
connection between BH spin and relativistic jets that are commonly
observed for both supermassive and stellar-mass BHs and that are so
prominent in the case of GRS1915 (e.g., Mirabel \& Rodr\'iguez 1999).
For many years, scientists have speculated that these jets are powered
by BH spin via a Penrose-like process associated with magnetic fields
(e.g., Blandford \& Znajek 1977; Hawley \& Balbus 2002; Meier 2003;
McKinney \& Gammie 2004).  However, these ideas will remain mere
speculation until sufficient data on BH spins have been amassed and the
models can be tested and confirmed.  This provides strong motivation for
measuring the spins of accreting BHs.

The strong evidence for natal spins -- particularly in the case of
GRS1915(\S6.2) -- is obviously of major significance in building
core-collapse models for SN and GRBs (Woosley 1993; MacFadyen \& Woosley
1999; Woosley \& Heger 2006).  For example, one of the greatest
uncertainties in GRB modeling is whether one can arrive at the core
collapse stage with sufficient angular momentum to make a disk around a
BH.  The spins of GRO~J1655--40 and 4U~1543--47 -- and especially
GRS1915 -- provide strong evidence for the high natal rotation rates of
BHs and thus provide strong support for the collapsar model of
``long-soft'' GRBs.

The continuing development of gravitational wave astronomy is central to
the exploration of BHs, and knowledge of BH spin is fundamentally
important to this effort.  To detect the faint coalescence signal for
two inspiralling BHs, one must compute the expected waveform and use it
to filter the data.  Our spin estimates for GRO~J1655-40 and 4U~1543--47
(R06) motivated the first such waveform computation that includes the
effects of spin (Campanelli et al. 2006), and our results reported here
for GRS1915 present a further challenge to the waveform modelers.

\subsection{An Observational Test of the Spin and Jet Models for
GRS~1915+105}

As mentioned in \S5.2, our fit results for the five low-luminosity
observations ($l < 0.3$; see \S4.2.1, Table~1) indicate that the
distance to GRS1915 is unlikely to be less than about 9--10~kpc.  This
result is based on the abrupt and dramatic rise in $\chi^{2}$ that
occurs for lesser distances.  For the nominal 14.0\msun~value of BH mass
and $D$ = 11.0, 10.5, 10.0, 9.5 and 9.0~kpc, the respective values of
$\chi_{\nu}^{2}$ for each low-luminosity observation are 0.6, 0.6, 2.6,
13.3, 43.4 (obs.\ no.\ 3); 0.5, 1.4, 2.7, 14.2, 43.8 (obs.\ no.\ 4);
0.6, 2.0, 6.3, 23.1, 263.6; (obs.\ no.\ 14); 1.0, 1.1, 5.6, 26.8, 77.1
(obs.\ no.\ 17); and 0.7, 0.7, 1.4, 12.1, 45.4 (obs.\ no.\ 20).  This
abrupt rise in $\chi^{2}$ indicates that we have reached the limit of
our table model ($a_* = 0.9999$) and that the fit is demanding
unphysical values of $a_* > 1$.  In Figure 18$a$, this distance lower
limit, which is a function of BH mass, is indicated by the long slant
line labeled ``spin model.''  For each assumed value of mass, and
hence inclination and distance (see \S5.2), the limiting value plotted
in Figure 18$a$ is an average result for the five low-luminosity points
at a 99\% level of confidence ($\Delta$$\chi^{2}$ = 6.6).  (The results
are very insensitive to the binary mass ratio, which we have held fixed
at its nominal observed value; Harlaftis \& Greiner 2004.)  To the right
of the vertical line labeled ``jet model,'' the intrinsic velocity of
the radio jet exceeds the velocity of light (\S5.2; Fender et al.\
1999).  The region below the nearly horizontal line is disallowed by the
jet model and the 1-$\sigma$ lower limit on the mass function (\S5.2;
Greiner et al.\ 2001).  Thus, taken together, the spin and jet models
plus the value of the mass function predict that the distance and BH
mass of GRS1915 lie within the triangular region shown in the figure.

Six model-dependent estimates of the distance to GRS1915 are summarized
in Figure 18$b$.  Some estimates disagree, others are very uncertain,
and none provides a convincing test of the constraints summarized in
Figure 18$a$.  We believe that it should be possible to obtain a
model-independent VLBA parallax distance that is precise to $\sim10$\%
and to reduce the uncertainty by a factor of two in the radial velocity
amplitude $K$ of the secondary, which would significantly improve the
accuracy of the mass function.  Such improvements in the observational
constraints will provide a powerful test of the spin and jet models for
GRS1915.


\section{Conclusions}

Using a rigorous and quantitative definition of the thermal state of a
black hole binary (\S2), we screened all the available {\it RXTE}~PCA
and {\it ASCA} GIS data and identified a total of 22 observations of
GRS~1915+105 that are free of QPOs and strong timing noise and for which
the thermal disk component of emission contributes $> 75$~\% of the
total 2--20~keV flux.  We then fitted the 22 disk-dominated spectra
using principally a model of a thin accretion disk in the Kerr metric
that includes all relativistic effects plus an advanced treatment of the
spectral hardening factor $f$ (\S4.2).  The spectral fitting of the 22
spectra was repeated a number of times using three different models for
the nonthermal tail component of emission and two different values of
the viscosity parameter (\S4).  The results for the key relativistic
parameters -- the spin $a_*$ and the mass accretion rate \mdot~-- were
shown to be quite independent of any details of the analysis and
insensitive to the uncertainties in the independently-determined input
parameters, namely, the mass, inclination and distance of the black hole
(\S5).

On theoretical grounds, we argue that the spin parameter can be
determined most reliably at lower luminosities (\S6.1, Appendix).  Our
relativistic disk model assumes a disk that is thin and torque-free at
its inner edge.  Higher luminosities are problematic because they likely
lead to disk thickening and nonzero torques near the ISCO.  Based on
theoretical arguments, we propose a limit on the disk thickness and a
corresponding limit on the disk luminosity, $L/L_{\rm Edd} < 0.3$, below
which one can obtain reliable estimates of the spin parameter.  Adopting
this criterion, we obtain our principal conclusion: GRS~1915+105 is a
rapidly-rotating BH with a lower limit on its spin parameter of $a_* >
0.98$.  Finally, we propose an observational test of our spin model.

\noindent



\acknowledgments 

We thank Keith Arnaud for help in implementing models in XSPEC and the
following people for helpful discussions and encouragement: Stan
Woosley, Alexandar Heger, Gerry Brown, Vicky Kalogera, Cole Miller, and
Paul Gorenstein.  We also thank an anonymous referee for helpful
comments and a thorough reading of our paper.  This research has made
use of data obtained from the High Energy Astrophysics Science Archive
Research Center (HEASARC), provided by NASA's Goddard Space Flight
Center. This work was supported in part by NASA grant NNG 05GB31G and
NSF grant AST 0307433.

\appendix

\section{Vertical Thickness of a Thin Accretion Disk around a Kerr Black Hole}

Following \citet{pag74}, we define the following functions for later use,
\begin{eqnarray}
	{\cal A} &=& 1+ a_*^2 x^{-4} + 2a_*^2 x^{-6} \;, \hspace{1cm}
	{\cal B} = 1+ a_* x^{-3} \;, \\
	{\cal C} &=& 1-3 x^{-2} + 2a_* x^{-3} \;, \hspace{1cm}
	{\cal D} = 1-2 x^{-2}+a_*^2 x^{-4} \;,
\end{eqnarray}
where $a_*\equiv a/c R_g$ is the dimensionless spin of the black hole
BH, $R_g\equiv GM/c^2$ is the gravitational radius of the BH of mass
$M$, and $x\equiv (R/R_g)^{1/2}$.  Note, ${\cal D}$ vanishes on the
horizon of the BH.  

On the equatorial plane of the BH, the lapse function and the angular
velocity of frame dragging are
\begin{eqnarray}
	\chi = \left(\frac{\cal D}{\cal A}\right)^{1/2} \;,
		\hspace{1cm}
	\omega = \frac{2 a_* R_g^2 c}{R^3 {\cal A}} \;.
\end{eqnarray}
The angular velocity of a thin Keplerian disk at radius $R$ is
\begin{eqnarray}
	\Omega_{\rm D} = \left(\frac{GM}{R^3}\right)^{1/2}
		\frac{1}{\cal B} \;,
\end{eqnarray}
and the rotational 3-velocity of the disk relative to the locally
nonrotating frame is
\begin{eqnarray}
	v_\phi = \frac{{\cal A}^{1/2}}{\chi}(\Omega_{\rm D}-\omega) R \;.
\end{eqnarray}
The 4-velocity of the disk particle is then 
\begin{eqnarray}
	U^a = \frac{\Gamma}{\chi}\left[\left(\frac{\partial}{\partial t}
		\right)^a + \Omega_{\rm D} \left(\frac{\partial}{\partial 
		\phi}\right)^a\right] \;,
	\label{u_a}
\end{eqnarray}
where $\Gamma = \left(1-v_\phi^2/c^2\right)^{-1/2} = {\cal B}/{\cal C}^{1/2}$ 
is the Lorentz factor. The 4-velocity satisfies the normalization condition
$U^a U_a = -1$.

The relative acceleration between two neighboring particles moving on
geodesics with a small separation vector $X^a$ is given by the
geodesic deviation equation \citep{wal84},
\begin{eqnarray}
	g^a = - R_{cbd}^{~~~a}X^bU^cU^d \;,
\end{eqnarray}
where $R_{cbd}^{~~~a}$ is the Riemann tensor of the spacetime and
$U^a$ is the four-velocity of the geodesic. The acceleration is
measured in the rest frame of the particles.  For a particle above the
equatorial plane at a small height $z$ and corotating with the disk,
we have $X^a = z e_z^a$, where $e_z^a$ is a normalized unit vector
orthogonal to the equatorial plane. Combining this with the $U^a$
given in equation~(\ref{u_a}) and the Riemann tensor of the Kerr
spacetime, we can calculate the relative acceleration,
\begin{eqnarray}
	g^a = - g_z e_z^a \;, \hspace{1cm}
	g_z = \xi \frac{GM z}{R^3} \;,
	\label{g_z}
\end{eqnarray}
where\footnote{Our result differs from eq. (5.7.2) of Novikov \&
Thorne (1973). After intensive examination, we believe that their
formula is incorrect.}
\begin{eqnarray}
	\xi = \frac{1}{\cal C}\left(1-4 a_* x^{-3}+3 a_*^2 x^{-4}
		\right) \;.
\end{eqnarray}

For a disk that is radiation-dominated (at least at the photosphere),
the equilibrium in the vertical direction is determined by
\begin{eqnarray}
	\frac{F\kappa}{c} \approx \left.g_z\right|_{z=H} \;, \label{fc}
\end{eqnarray}
where $F=F(R)$ is the radiation flux density of the disk (measured by
an observer corotating with the disk) and $\kappa$ is the disk
opacity.
The flux density $F$ has been derived by \citet{pag74} and is given by
\begin{eqnarray}
	F = \frac{3GM\dot{M}}{8\pi R^3} f_0 \;,
\end{eqnarray}
where $f_0 = (2R^2/3R_g) f$ and the expression for $f$ is given in
equation~(15n) of \citet{pag74}. Our choice of $f_0$ instead of $f$ is
based on the fact that, unlike $f$, $f_0$ is dimensionless. Note that
at $R = R_{\rm ISCO}$ (the innermost stable circular orbit) we have
$f_0 = 0$, and that as $R\rightarrow\infty$ we have $f_0\rightarrow
1$.

By equations~(\ref{g_z})--(\ref{fc}), the scale-height of the disk is
\begin{eqnarray}
	\frac{H}{R} \approx \frac{3\kappa\dot{M}}{8\pi Rc}
		\frac{f_0}{\xi} \;.
	\label{hr0}
\end{eqnarray}
Following the analysis of the Newtonian case (\S~6.1), we define the
Eddington luminosity by
\begin{eqnarray}
	L_{\rm Edd} = \frac{4\pi GM c}{\kappa} = \varepsilon 
		\dot{M}_{\rm Edd} c^2 \;, 
		\hspace{1cm} \mbox{i.e.,}\hspace{1cm}
	\dot{M}_{\rm Edd} = \frac{4\pi c R_g}{\kappa \varepsilon} \;,
\end{eqnarray}
except that here $\varepsilon = \varepsilon(a_*)$ is the radiative
efficiency of the relativistic disk (see Page \& Thorne 1974). With
the above definition of $\dot{M}_{\rm Edd}$, we have $L/L_{\rm Edd} =
\dot{M}/\dot{M}_{\rm Edd} \equiv \dot{m}$, where $L$ is the luminosity
of the disk.  Then, equation~(\ref{hr0}) can be recast into
\begin{eqnarray}
	\frac{H}{R} \approx \frac{3\dot{m}}{2\varepsilon}\frac{f_0}
		{x^2\xi} \;.
	\label{hr}
\end{eqnarray}
It can be shown that this expression for $H/R$ simplifies to
equation~(\ref{HoverR}) in the Newtonian limit.  Note that $H/R$ does
not depend on the value of the opacity $\kappa$.

Since the ratio $H/R=0$ at $R=R_{\rm ISCO}$ and also as
$R\rightarrow\infty$, it must have a maximum at some finite $R>R_{\rm
ISCO}$.  It turns out that for any given value of $\dot{m}$, the
maximum value of $H/R$ is very insensitive to variation in $a_*$
(though the radius at which this maximum is reached varies by a large
factor). Examples of $H/R$ as a function of the disk radius are shown
in Figure~17 for two choices of the BH spin, $a_*=0, ~0.998$.  Notice
how the two sets of curves agree very closely as far as their maxima
are concerned.  Therefore, regardless of the value of $a_*$, if we
wish to have $(H/R)_{\rm max} \lesssim 0.1$, we require the
dimensionless disk luminosity $l \equiv L/L_{\rm Edd}$ to be $\lesssim
0.3$.

\clearpage

\figcaption[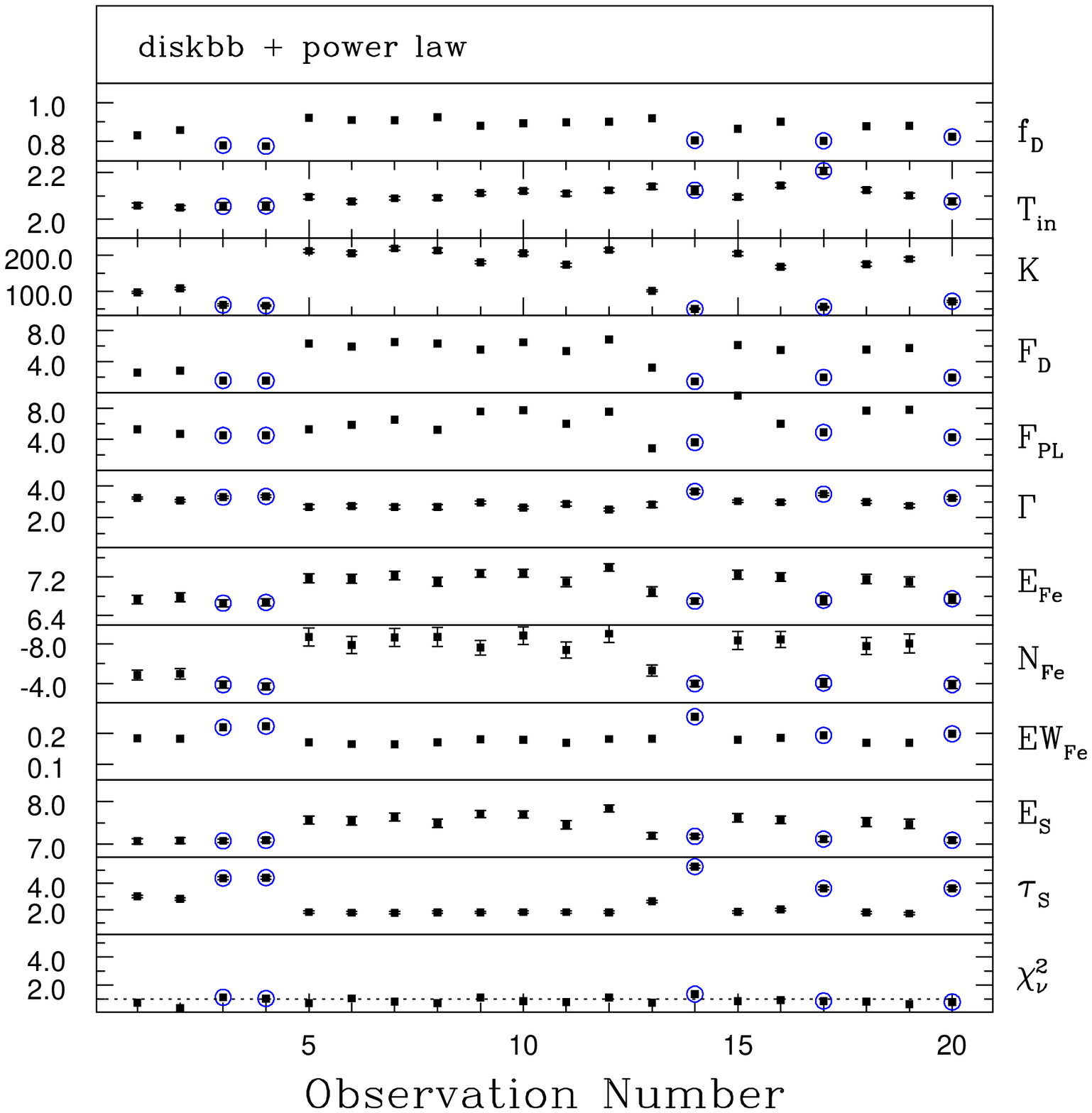] {Detailed results from fitting all 20 {\it RXTE}
observations of GRS1915 in the thermal state with a model consisting of
{\it diskbb}, a power-law, a Gaussian absorption line and a smedge
component over the energy range 3--25 keV (44 degrees of freedom).  The
data points circled in blue correspond to the crucial low-luminosity
observations (\S5, \S6.1, Appendix).  The horizontal dashed line in the
bottom panel is drawn at $\chi_{\nu}^{2} = 1$.  From top to bottom, the
panels show: the ratio of the disk to total flux $f_D$, the two {\it
diskbb} fitting parameters, the disk inner temperature $T_{\rm in}$
(keV) and the normalization constant $K$, the disk flux $F_D$
(10$^{-7}$~\ergcm) and the power-law flux $F_{\rm PL}$
(10$^{-8}$~\ergcm), the power-law photon index $\Gamma$, the central
energy of the Gaussian absorption line $E_{\rm Fe}$ (keV), the intensity
of the line $N_{\rm Fe}$ (photons~cm$^{-2}$~s$^{-1}$ times 100) and the
equivalent width of the line ${\rm EW_{Fe}}$ (keV), the smedge edge
energy $E_S$ (keV) and the smedge optical depth $\tau_S$, and finally
the value of reduced chi-square.  See \S4.1 for further details.}

\figcaption[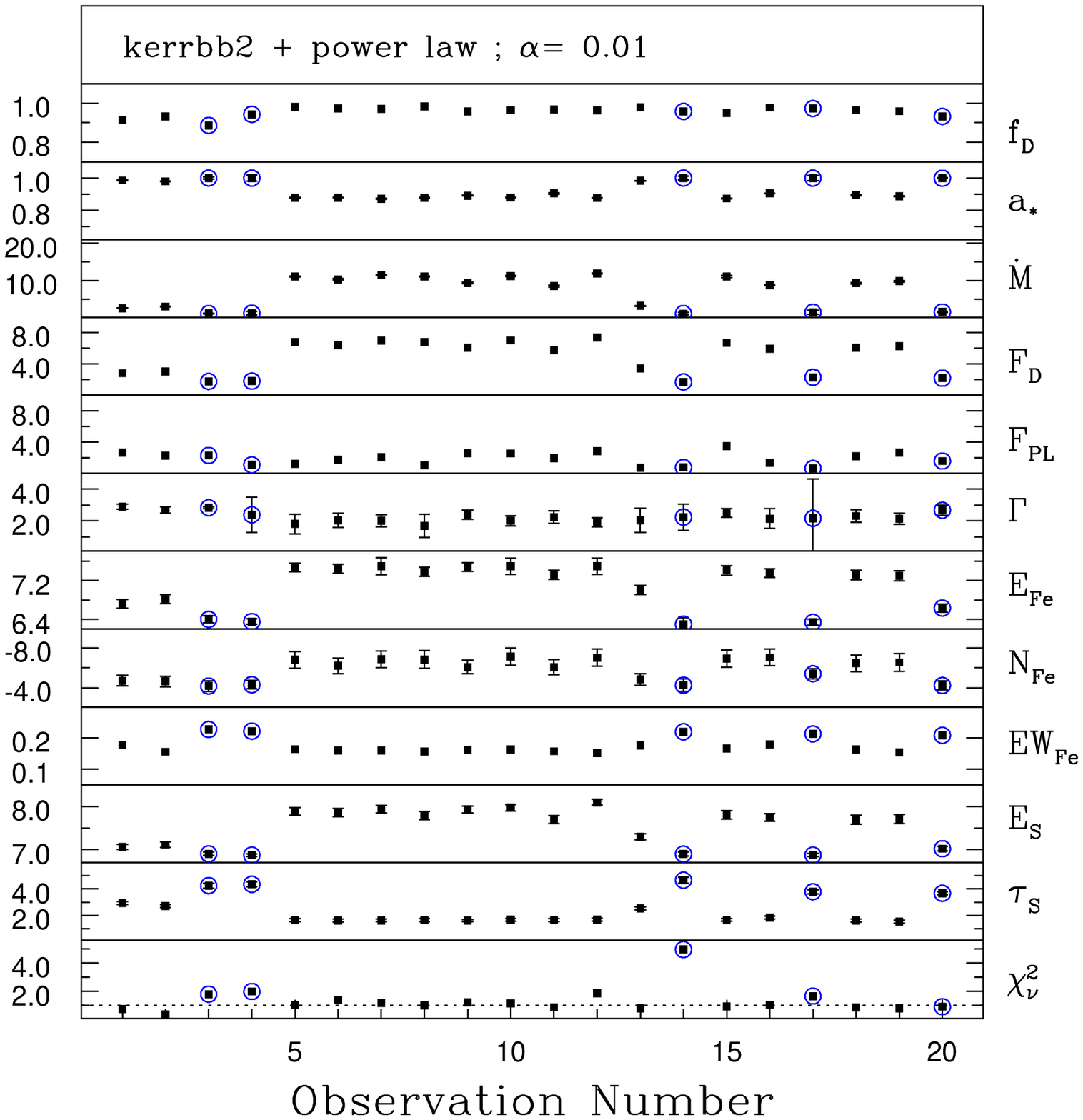] {Analogous to Figure 1, but with the
non-relativistic disk model {\it diskbb} replaced by our relativistic
disk model {\it kerrbb2} (44 dof).  The fits were done for a viscosity
parameter $\alpha=0.01$.  This figure is identical in structure with
Figure~1 except that $T_{\rm in}$ and $K$ are here replaced by two
parameters of {\it kerrbb2}, namely, the BH spin parameter $a_*$ and the
mass accretion rate \mdot\ (${\rm 10^{18} ~g\,s^{-1}}$).}

\figcaption[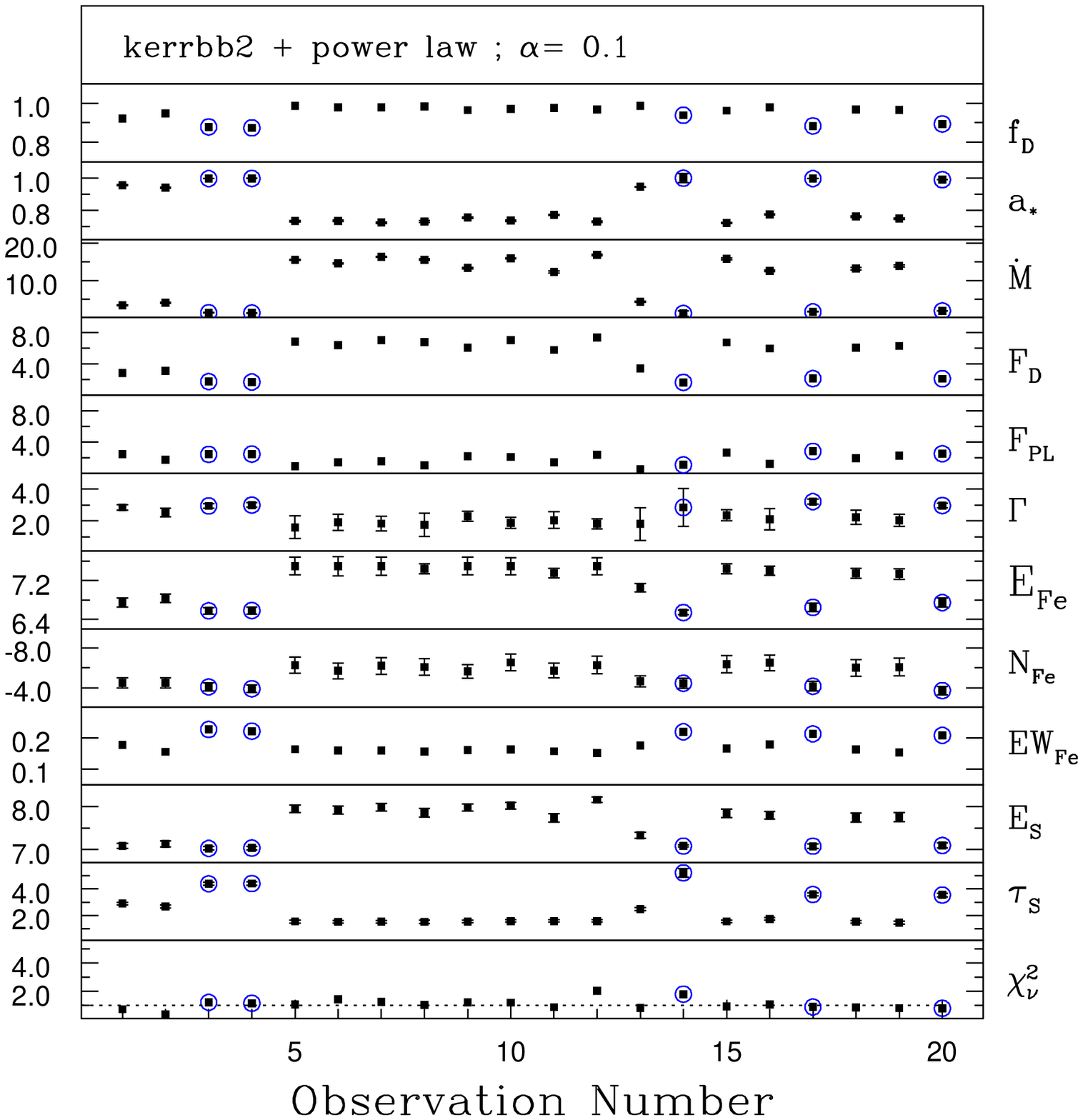] {This figure is identical to Figure~2, except that
the fits were computed for a viscosity parameter $\alpha = 0.1$.}

\figcaption[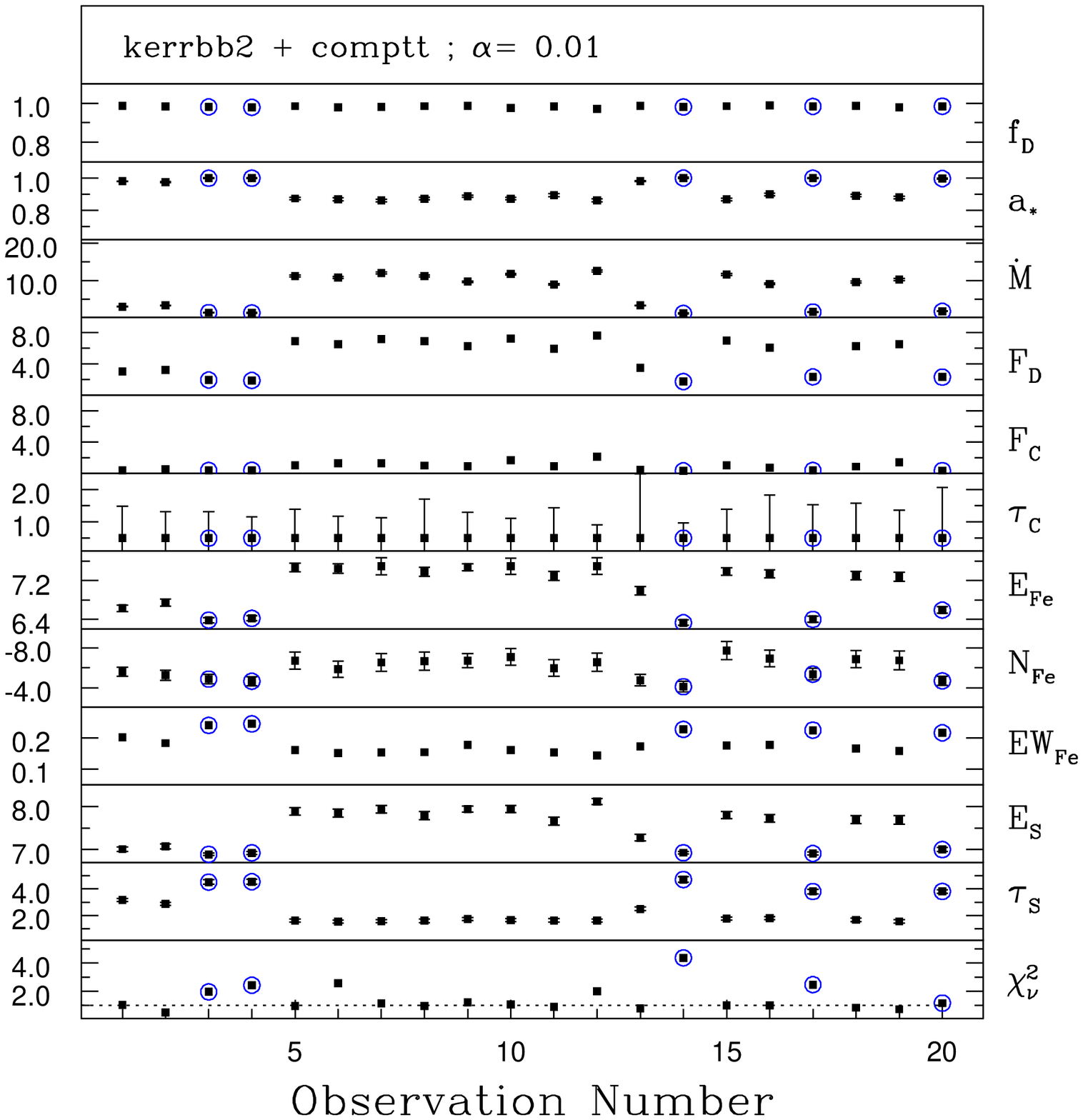] {Results of fitting the 20 {\it RXTE} observations
of GRS1915 in the thermal state with a model consisting of {\it
kerrbb2}, a thermal Comptonization component {\it comptt}, a Gaussian
absorption line and a smedge component (43 dof).  The panels are the
same as in Figures 2 and 3 except that $F_{\rm PL}$ and $\Gamma$ are
replaced by the flux in the {\it comptt} component $F_C$ and the optical
depth of the Comptonizing corona $\tau_C$.  See \S4.2.2 for other
details.}

\figcaption[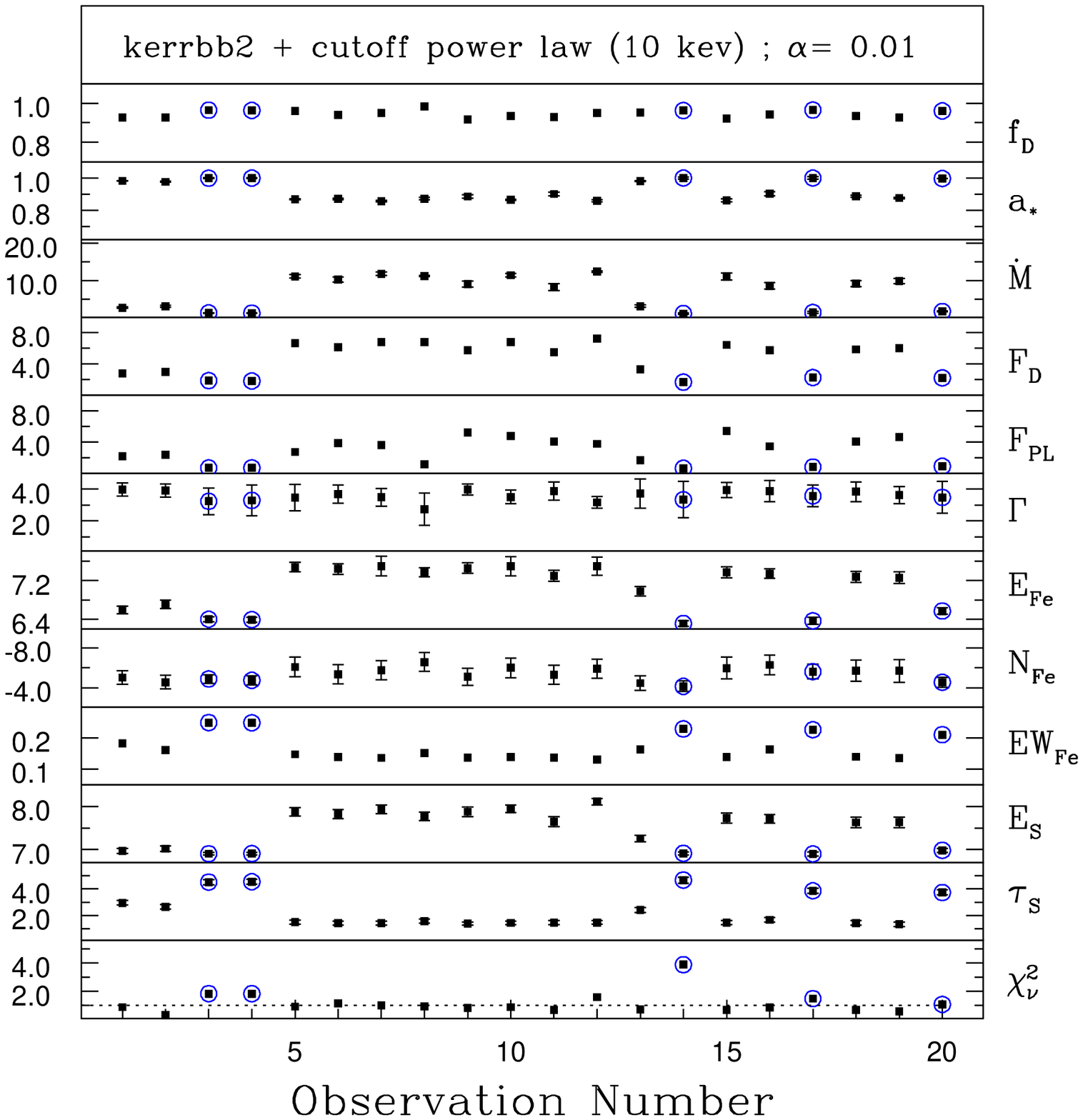] {Results of fitting the 20 {\it RXTE} observations
of GRS1915 in the thermal state with a model consisting of {\it
kerrbb2}, a cutoff power-law component {\it expabs*power}, a Gaussian
absorption line and a smedge component (43 dof).  See \S4.2.3 for other
details.}

\figcaption[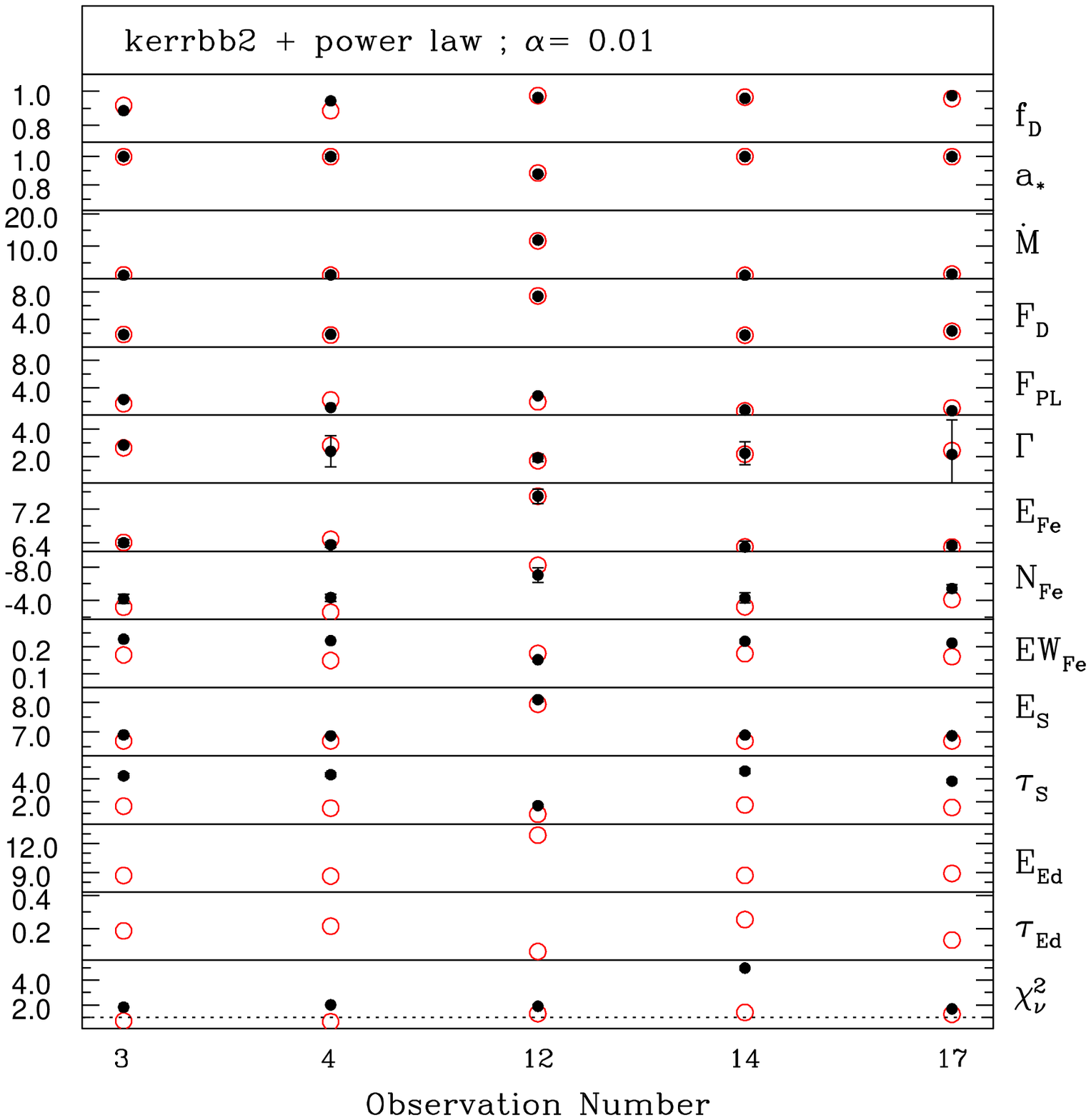] {Fit results for a model including a sharp
absorption edge for the five observations in Figures~2--5 that have
$\chi_{\nu}^{2} \gtrsim 1.5$.  The small black data points with error
bars are identical to those plotted in Figure~2, and the results of
including the sharp edge in the fits are plotted as open red circles.
Note the pair of panels near the bottom displaying the parameters of the
edge component.}

\figcaption[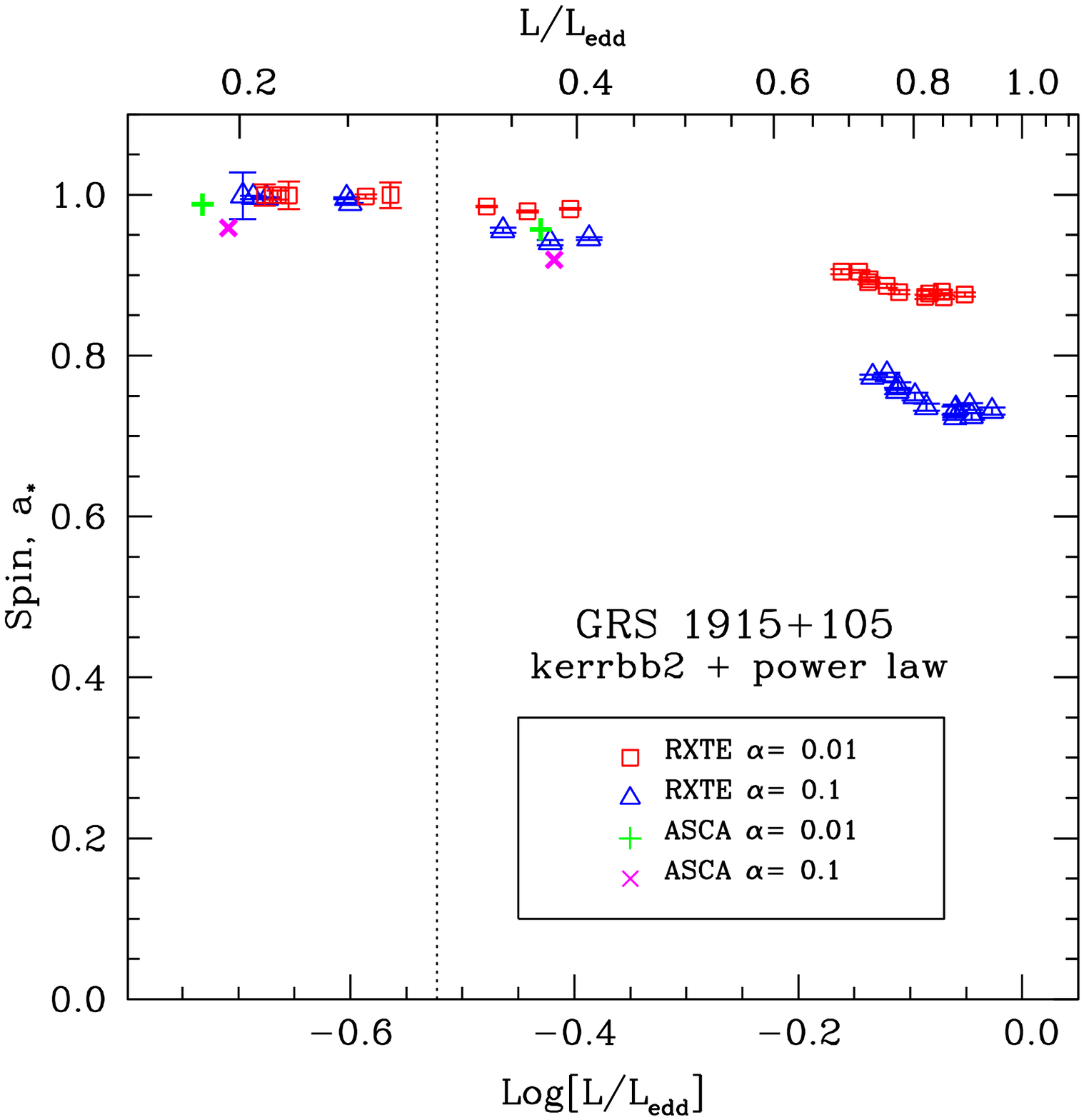] {Spin parameter $a_*$ versus the Eddington-scaled
luminosity $L/L_{\rm Edd}$ for all 22 {\it RXTE} and {\it ASCA}
observations of GRS1915 in the thermal state for two values of the
viscosity parameter $\alpha$.  The tail emission is modeled as a simple
power law.  For reasons discussed in \S~6.1 and the Appendix, the
results are most trustworthy for $L/L_{\rm Edd} \lesssim 0.3$; this
limit is indicated here and below by the vertical dotted line.  Data in
this regime consistently give a very high estimate of the spin parameter
of GRS1915, $a_* \to 1$, independent of $\alpha$ or any other details.}

\figcaption[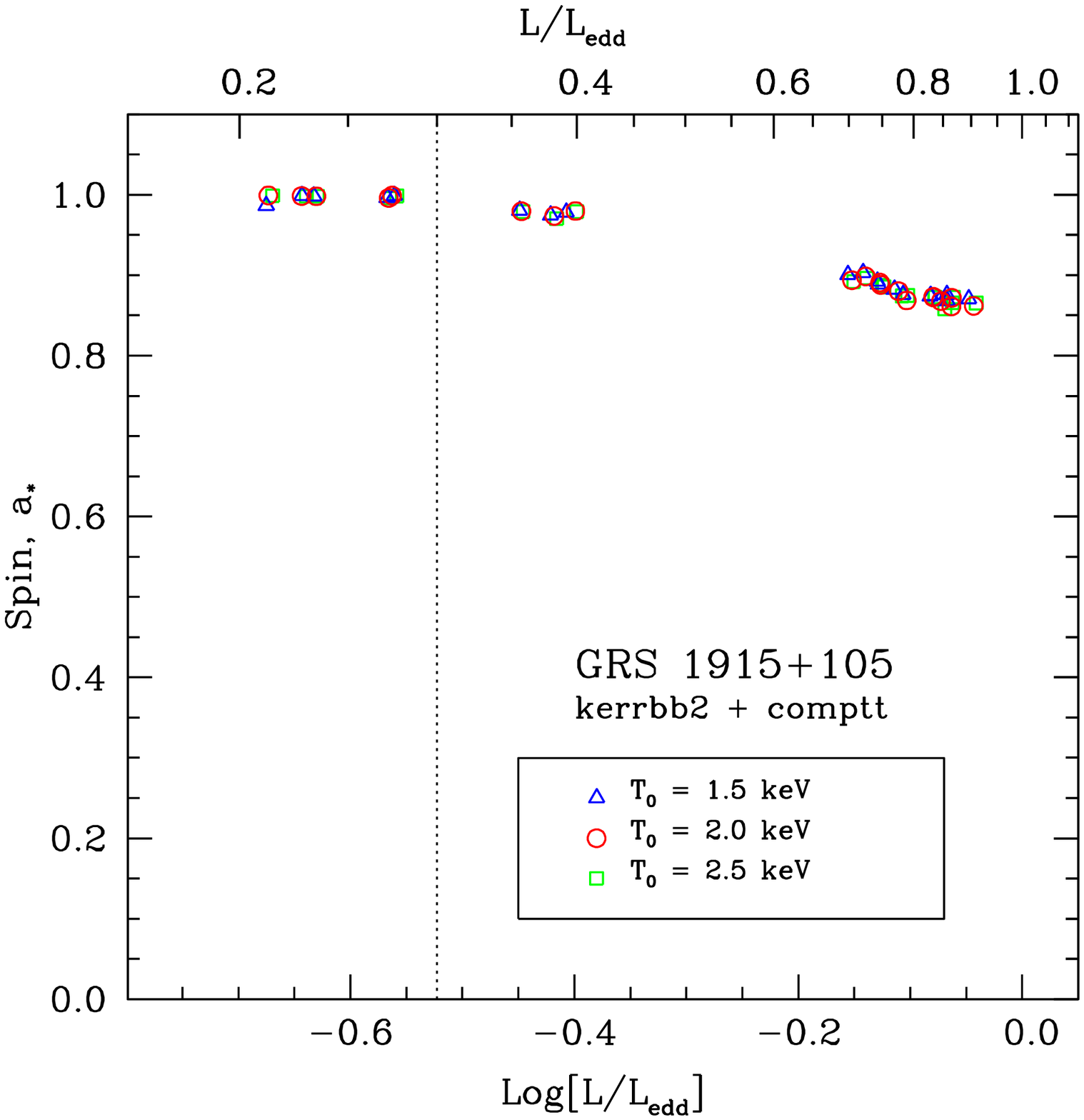] {Same as Figure~7 except that the tail emission is
modeled as a Comptonized plasma and only the results for $\alpha = 0.01$
are shown.  Results are displayed for three values of $T_{\rm 0}$, the
temperature of the seed photons.}

\figcaption[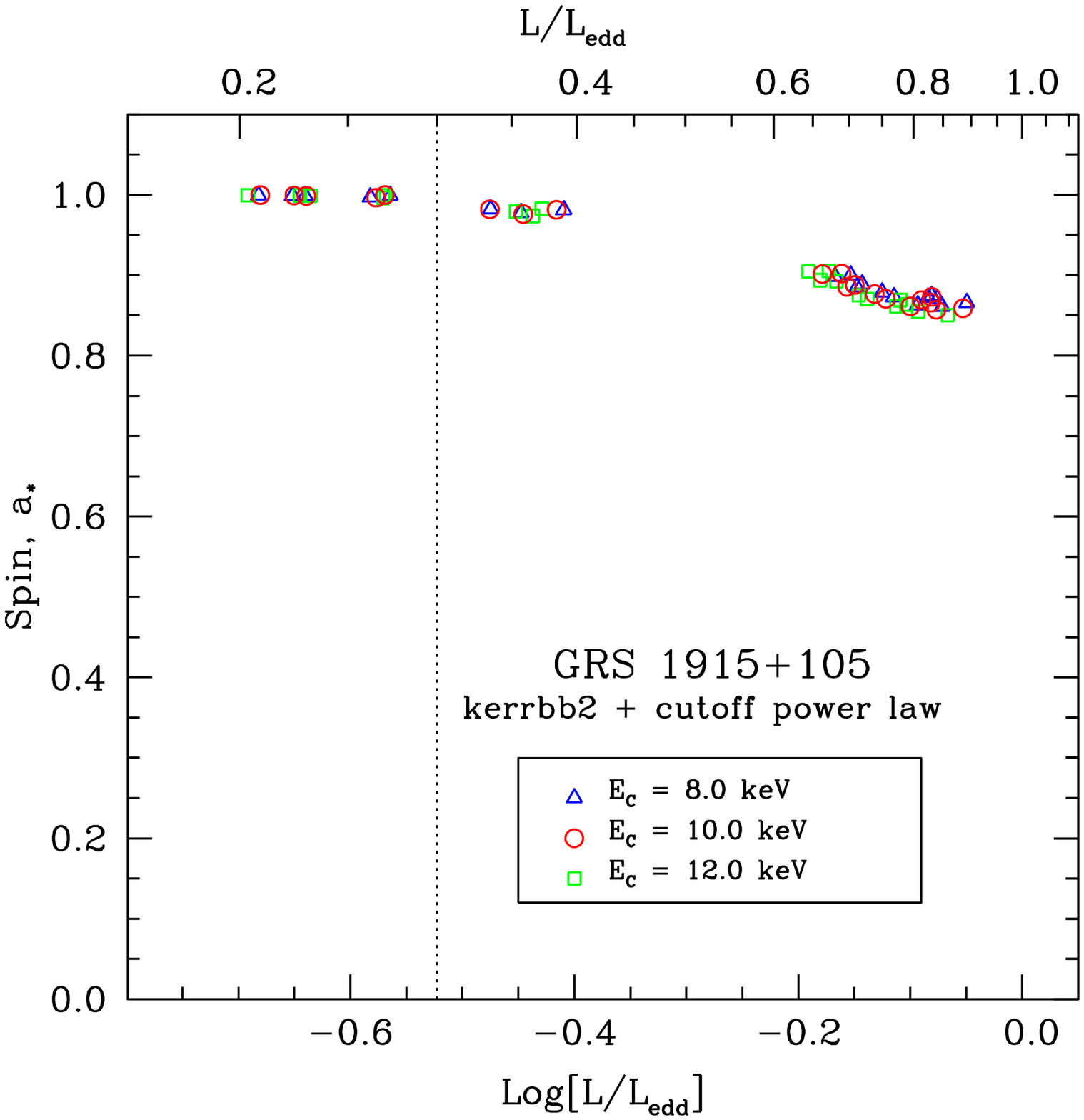] {Same as Figure~7 except that the tail emission is
modeled as a cutoff PL, and $\alpha = 0.01$ only. The results are shown
for three values of the cutoff energy $E_{\rm c}$.}

\figcaption[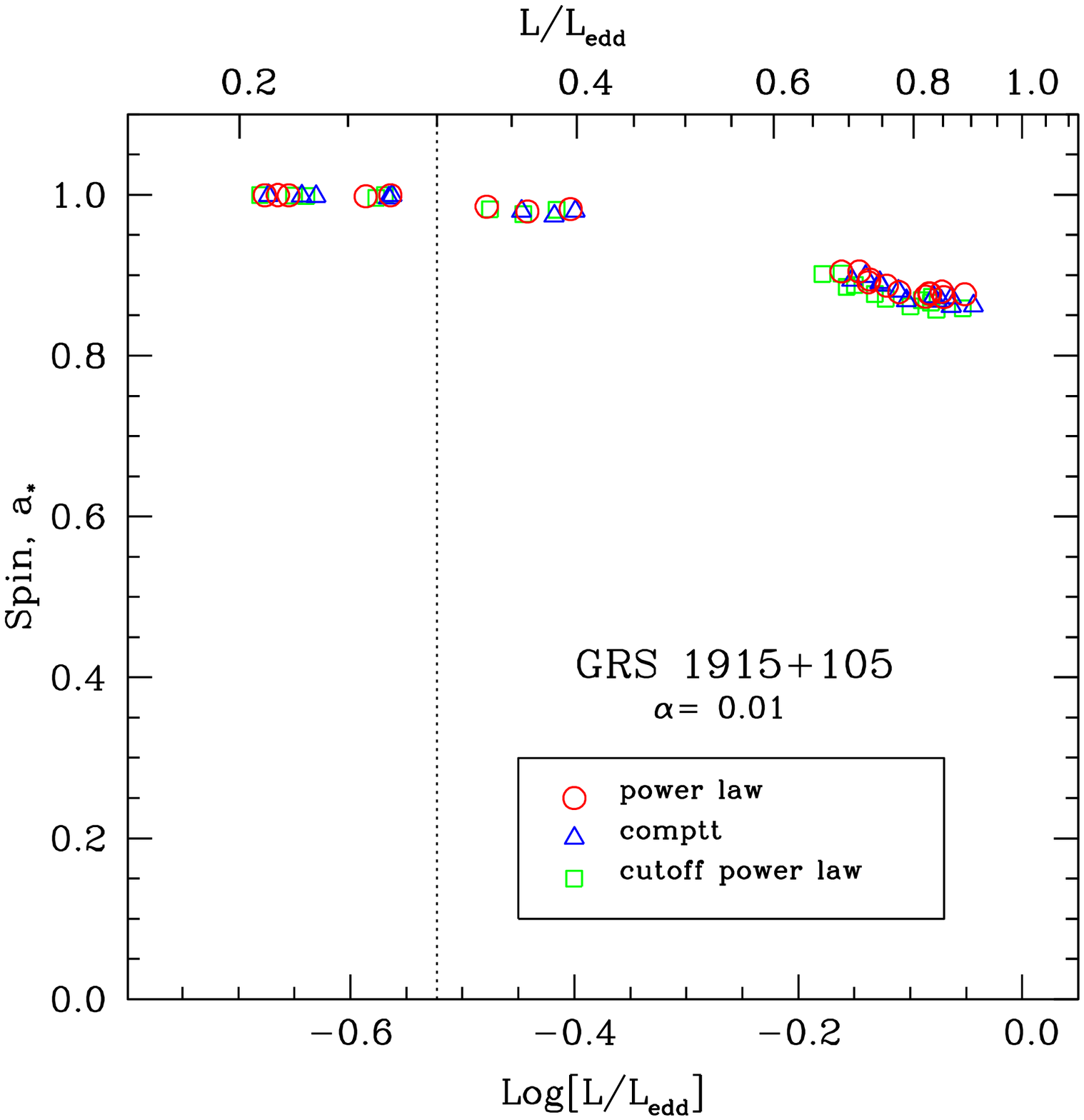] {Direct comparison of the results displayed in
Figures 7--9 for the three different tail models, for $\alpha = 0.01$
only.  Note how very similar the results are, which shows that the
results are not sensitive to the details of the spectral model used to
fit the high-energy tail component in the spectrum.}

\figcaption[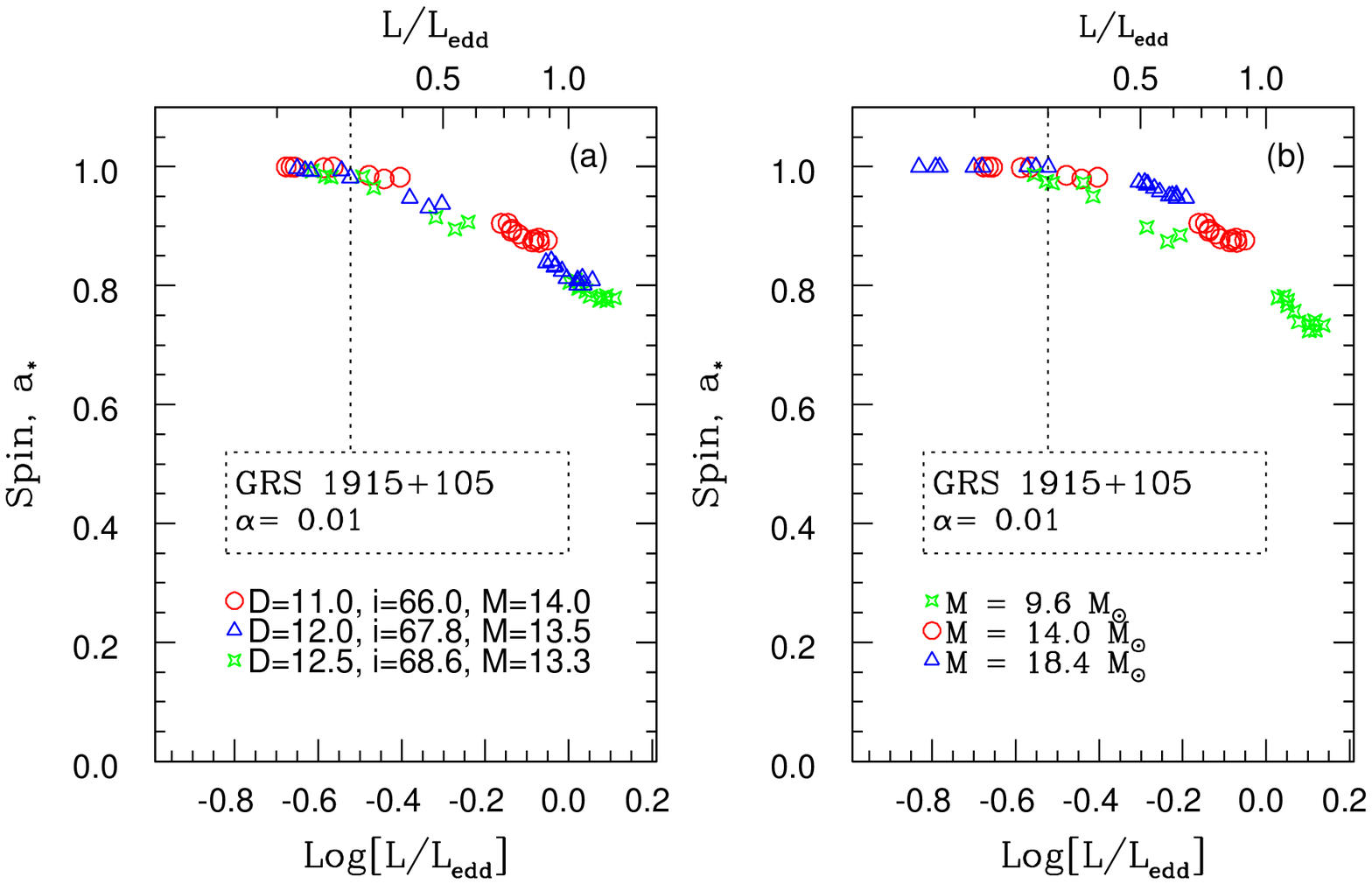] {($a$) Effects on the spin estimate of GRS1915 as a
result of varying the distance $D$ to the source over the range
11.0--12.5 kpc.  The mass of the BH $M$ and the inclination $i$ are
correlated with $D$, as explained in \S5.2.  The results for $D = 9$~kpc
and $D = 10$~kpc are not shown for reasons that are given in \S5.2.
($b$) Effects of varying the BH mass $M$ over its allowed
range, keeping $D$ fixed at 11.0 kpc and $i$ fixed at $66.0^o$.}

\figcaption[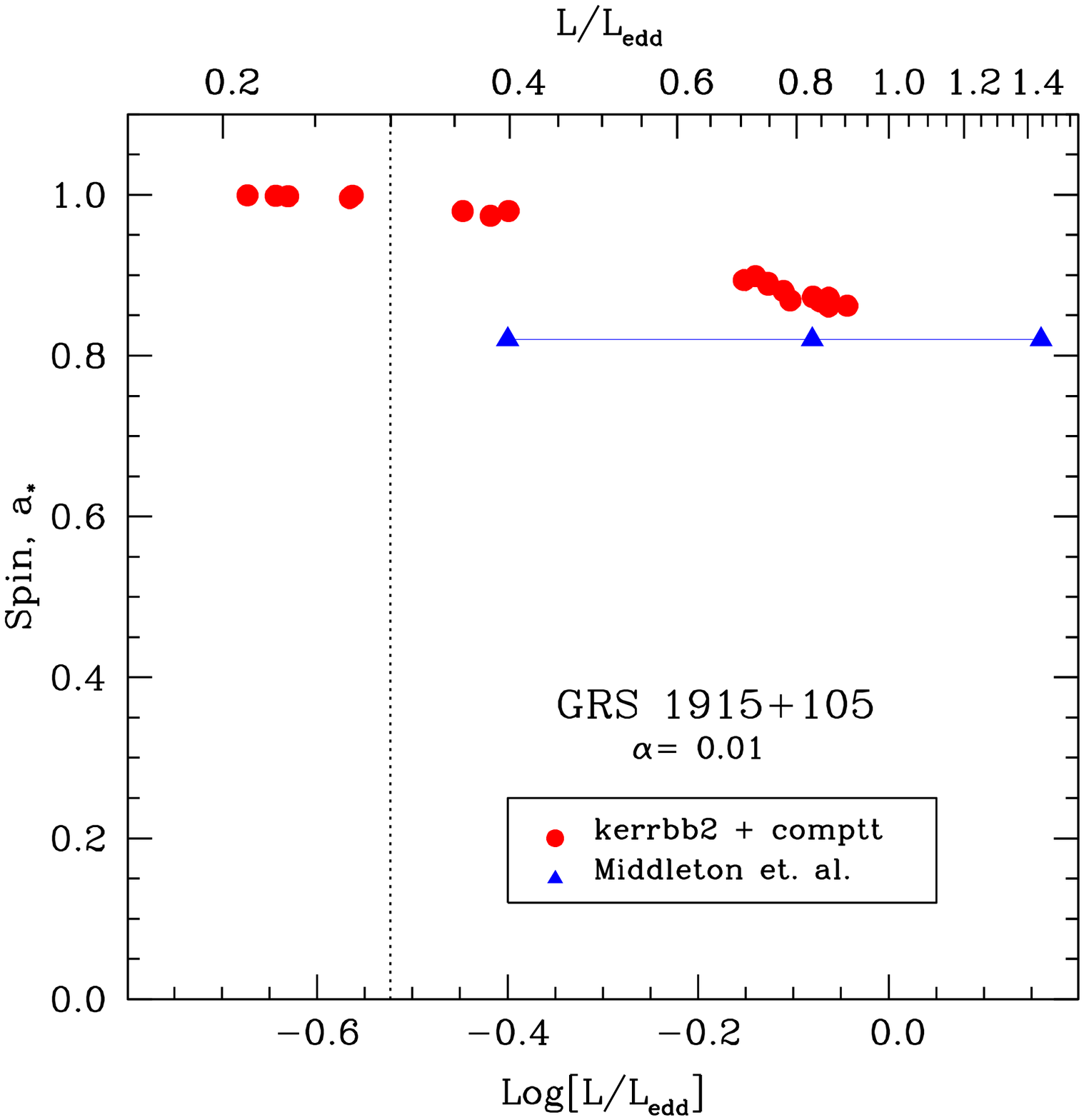]{The single spin estimate obtained by M06, which
is here referred to a distance $D = 11$~kpc, is indicated by the three
blue triangles that are connected by a dashed line.  Our results,
which are based on the {\it comptt} tail model for 
$T_{\rm 0} = 2.0$~keV and $\alpha = 0.01$,
are shown as red circles (see Fig.\ 8).}

\figcaption[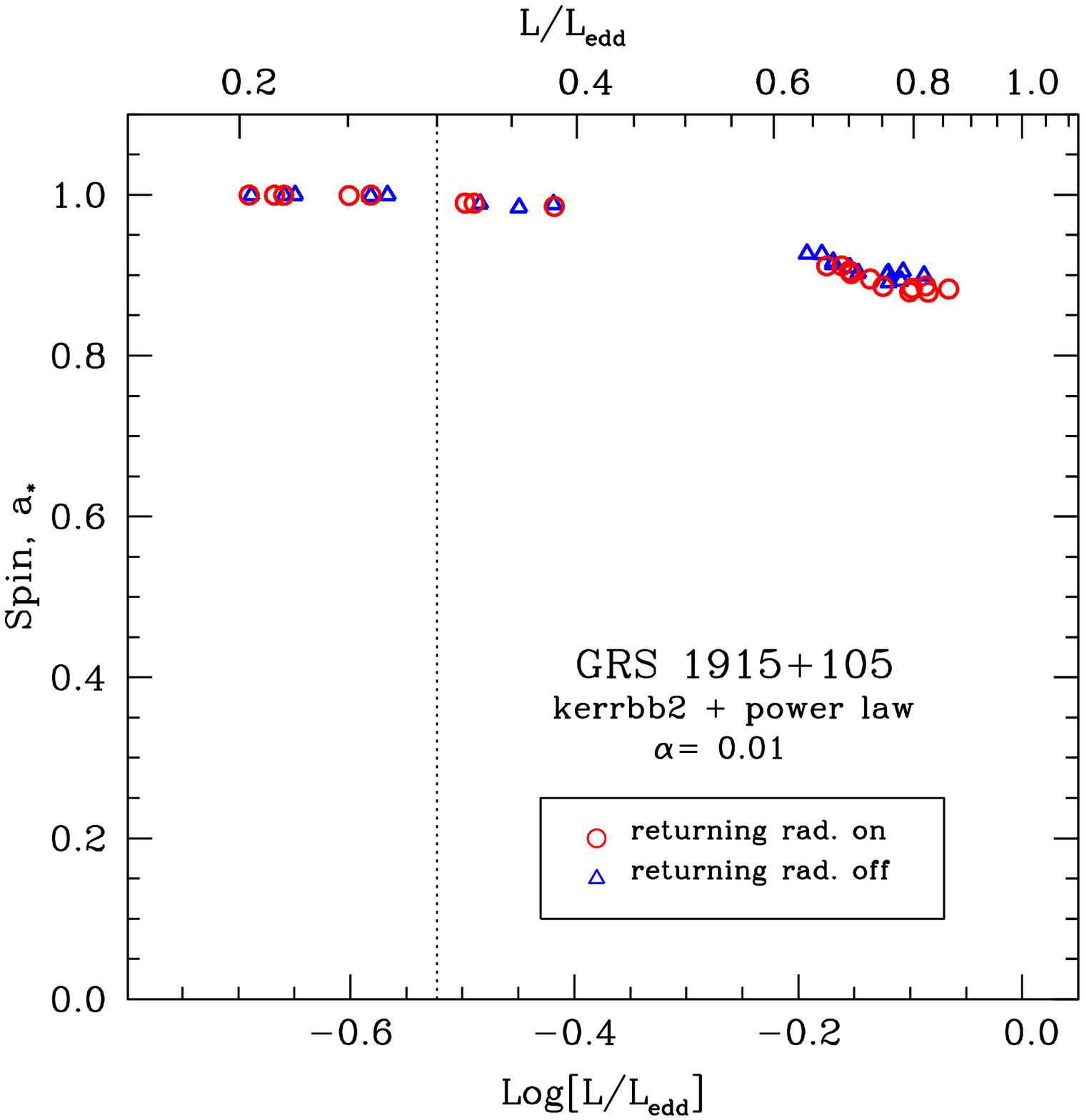]{Illustrates the effect of including the returning
radiation in the model.  The primary effect is to shift the estimated
Eddington-scaled luminosities to higher values.  There is very little
effect on the estimates of BH spin $a_*$.}

\figcaption[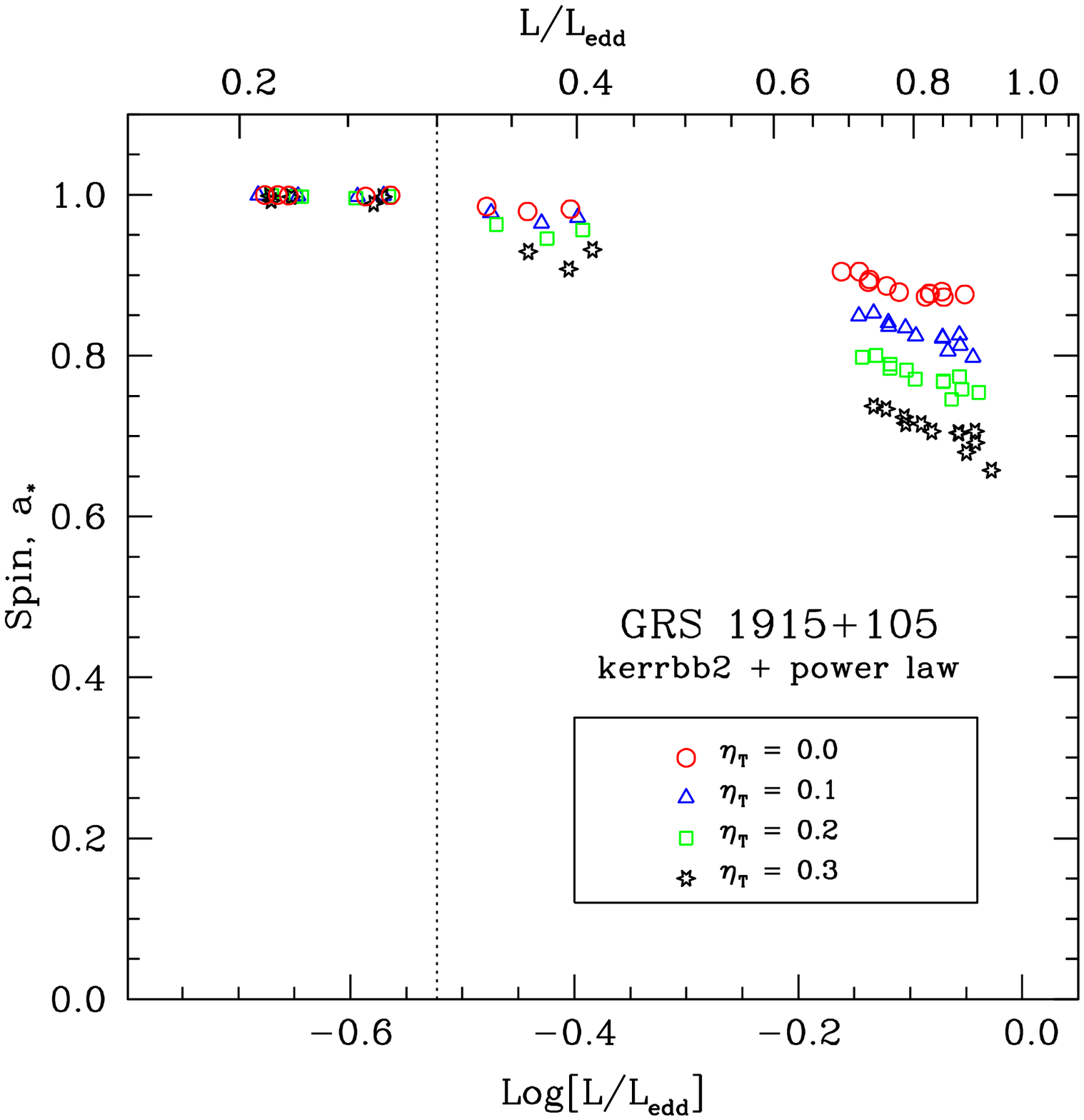]{Illustrates the effect of including a nonzero
torque at the inner edge of the disk.}

\figcaption[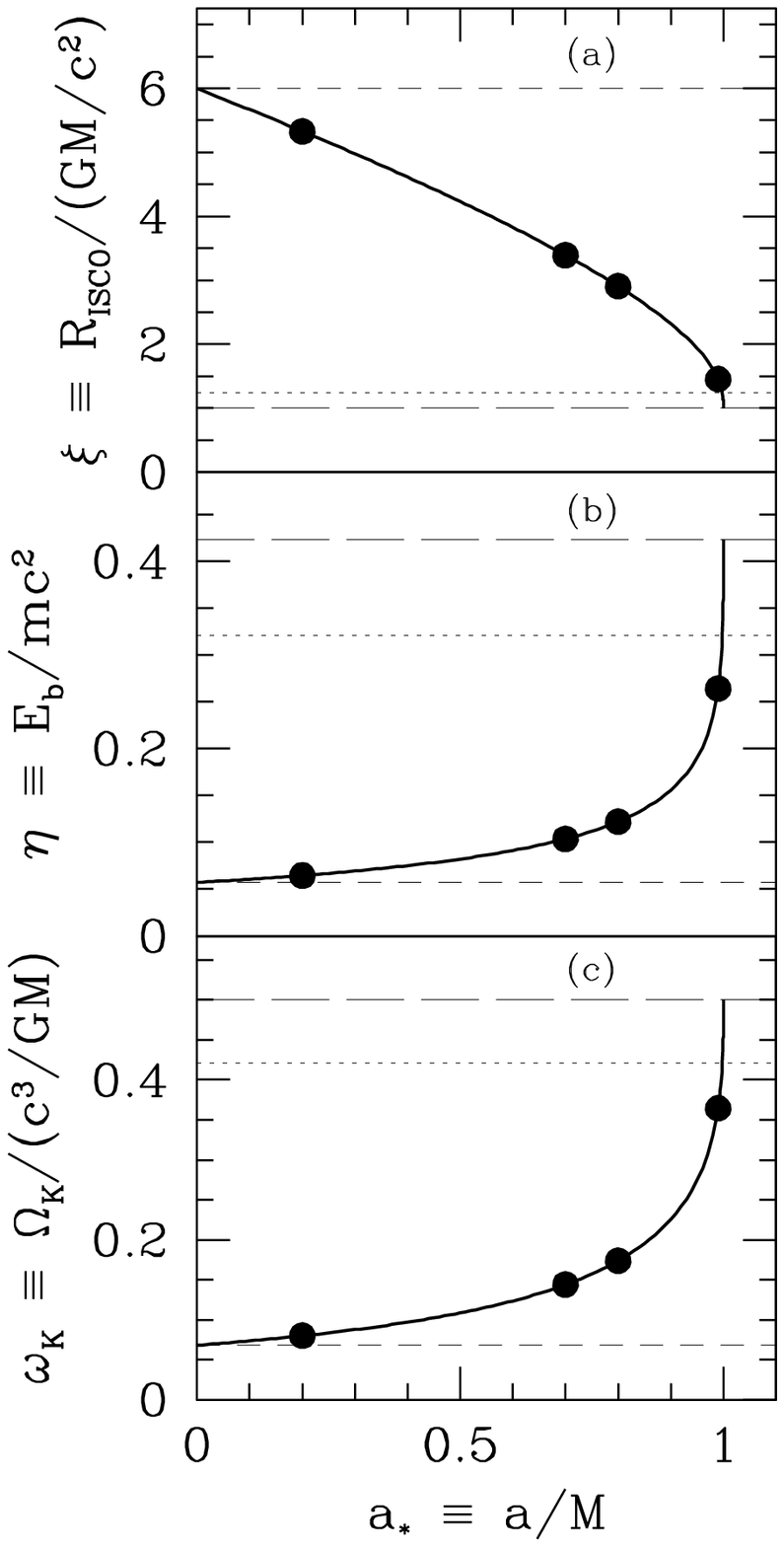]{The behavior of three dimensionless quantities that
depend only on the BH spin parameter: (a) The radius of the ISCO in
gravitational units, (b) the specific binding energy at the ISCO, and
(c) the Keplerian orbital frequency at the ISCO.  The filled data points
correspond to nominal estimates of the spins of the four BHs (see
Table~4): from left to right, LMC~X-3 ($a_*=0.20$), GRO~J1655--40
($a_*=0.70$), 4U~1543--47 ($a_*=0.80$), and GRS1915 ($a_*=0.99$).  The
horizontal lines in each panel indicate the values of each of the three
quantities in question that correspond to the following key values of
spin: $a_* = 0$ (short-dashed line), $a_* = 1$ (long-dashed line), and
$a_* = 0.998$ (dotted line; Thorne 1974).}

\figcaption[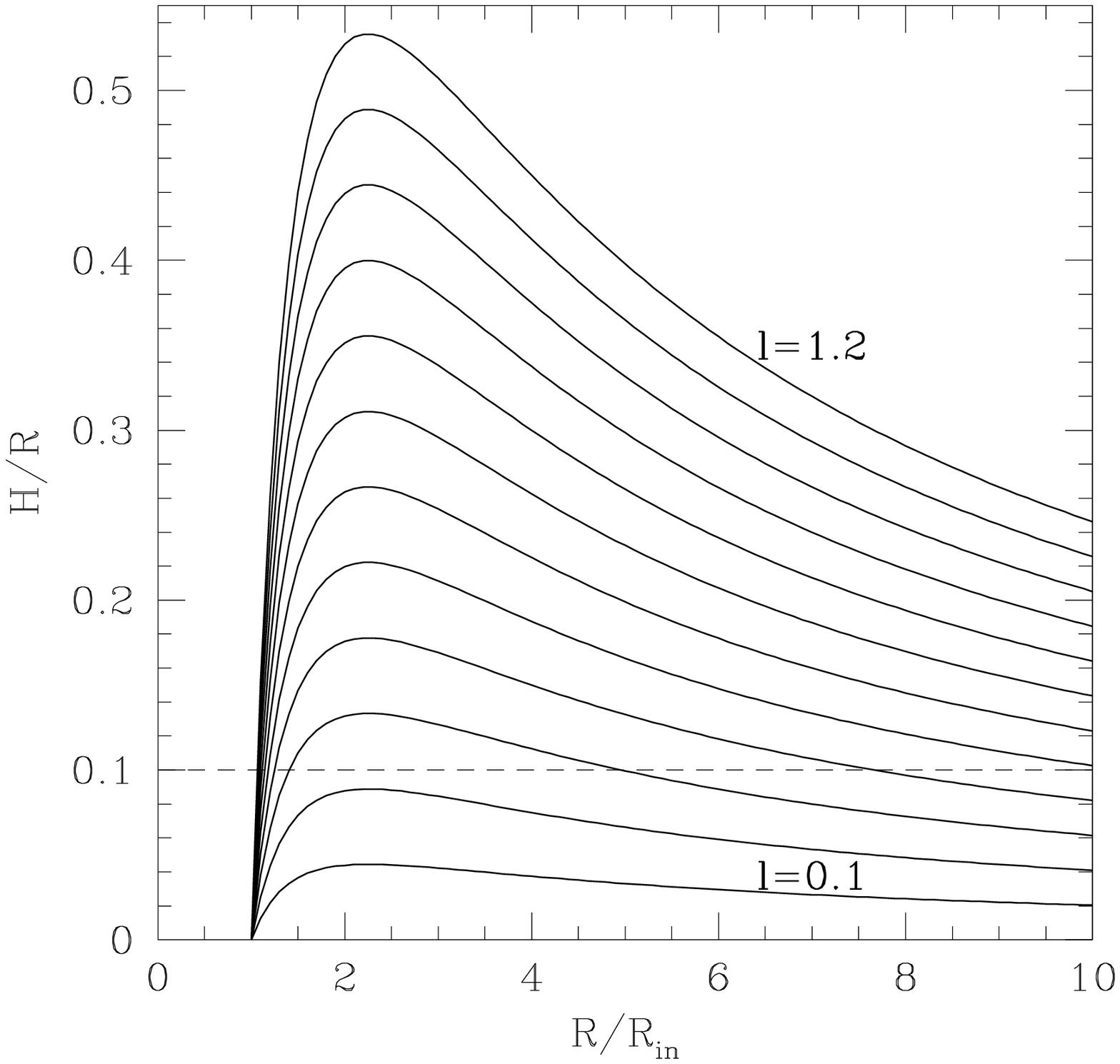]{Ratio of disk thickness $H$ to radius $R$, plotted
against $R/R_{\rm in}$, for Eddington-scaled values of luminosity
$l=\dot{m}$ in steps of $\Delta l = 0.1$ (from $l=0.1$ to $l=1.2$
upward).  The results are for a Newtonian disk in which $R_{\rm in}$ is
the radius of the inner edge.  The horizontal dashed line corresponds to
$H/R=0.1$.  (See Fig.\ 17 for the relativistic case.)}

\figcaption[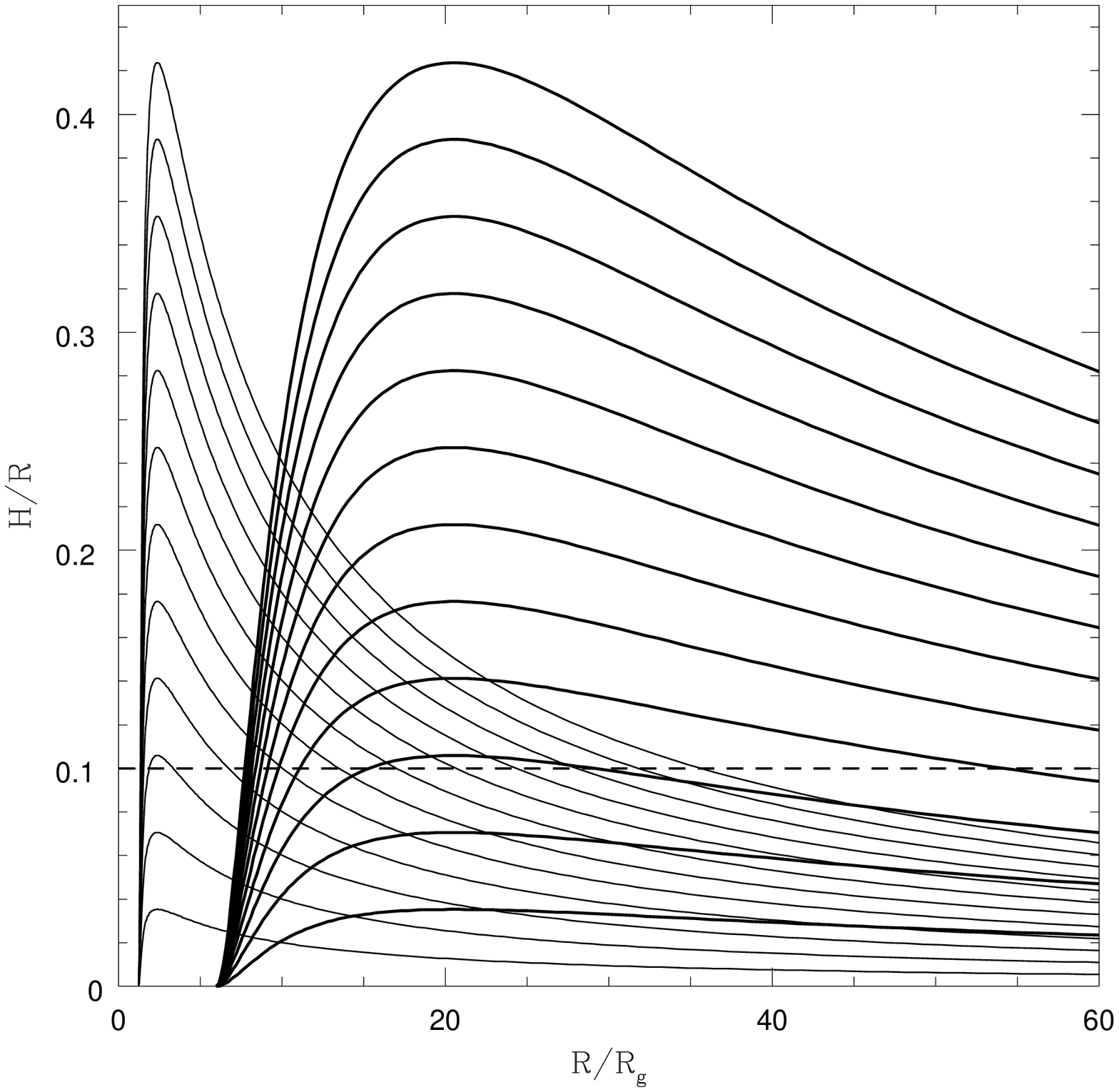]{Ratio of disk thickness $H$ to radius $R$ for a
relativistic disk around a Kerr black hole, plotted against $R/R_g$, for
Eddington-scaled values of luminosity $l=\dot{m}$ in steps of $\Delta l
= 0.1$ (from $l=0.1$ to $l=1.2$ upward).  The inner radius of the disk
is at the innermost stable circular orbit. The thick lines correspond to
a non-rotating black hole ($a_*=0$) and the thin lines to a maximally
rotating black hole ($a_*=0.998$). The horizontal dashed line
corresponds to $H/R=0.1$.  It is anticipated that the disk spectral
models employed in this paper ({\it diskbb, kerrbb2, bhspec}) are most
reliable when $H/R\lesssim0.1$, which corresponds to the luminosity
limit $l \lesssim 0.3$.  (See Fig.\ 16 for the Newtonian case.)}

\figcaption[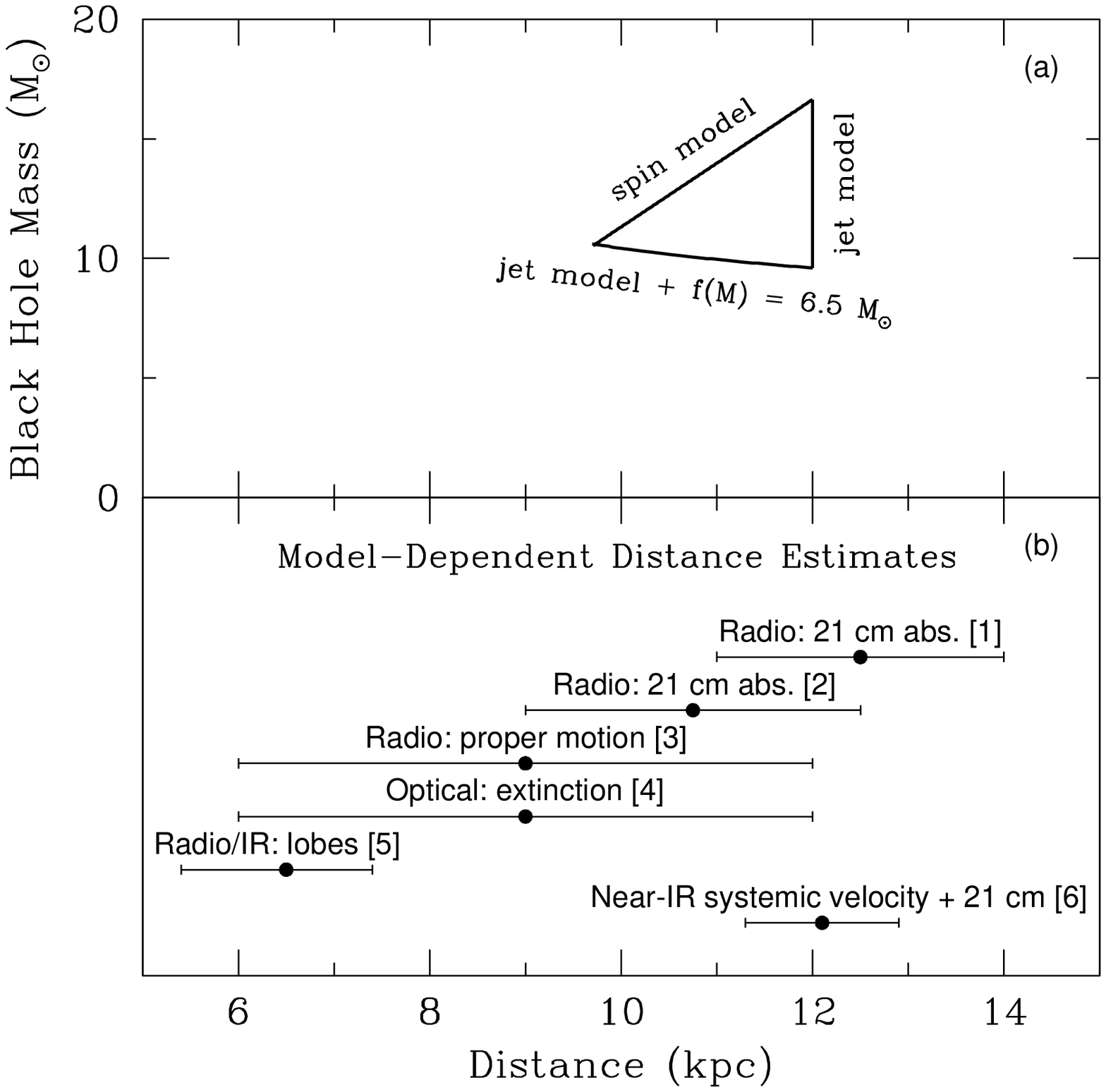]{($a$) The allowed values of BH mass and distance
for GRS1915 fall within the triangular region indicated (see text).
($b$) A summary of model-dependent distance estimates for GRS1915.  The
two relatively precise and disparate estimates at the bottom of the
figure require comment: The one labeled ``Radio/IR: lobes'' is based on
identifying a pair of extended IRAS sources as the regions where the
jets of GRS1915 impact the ISM (Kaiser et al.\ 2004).  The other
estimate labeled ``Near-IR systemic velocity + 21 cm'' is based on a
systemic velocity of $\gamma = -3 \pm 10$ km~s$^{-1}$ and the Galactic
rotation curve (Greiner et al.\ 2001).  This latter estimate ignores the
potentially sizable and unknown uncertainty associated with a possible
peculiar component of radial velocity as well as any kick velocity that
may have been imparted to the system during the formation of the BH
(e.g., Jonker \& Nelemans 2004).  References: (1) Rodr\'iguez et al.\
1995; (2) Dhawan et al.\ 2000a; (3) Dhawan et al.\ 2000b; (4) Chapuis \&
Corbel 2004; (5) Kaiser et al.\ 2004; (6) Greiner et al.\ 2001.}

\clearpage
\begin{figure}
\figurenum{1}
\plotone{f1.eps}
\caption{ }
\end{figure}

\clearpage
\begin{figure}
\figurenum{2}
\plotone{f2.eps}
\caption{ }
\end{figure}

\clearpage
\begin{figure}
\figurenum{3}
\plotone{f3.eps}
\caption{ }
\end{figure}

\clearpage
\begin{figure}
\figurenum{4}
\plotone{f4.eps}
\caption{ }
\end{figure}

\clearpage
\begin{figure}
\figurenum{5}
\plotone{f5.eps}
\caption{ }
\end{figure}

\clearpage
\begin{figure}
\figurenum{6}
\plotone{f6.eps}
\caption{ }
\end{figure}

\clearpage
\begin{figure}
\figurenum{7}
\plotone{f7.eps}
\caption{ }
\end{figure}

\clearpage
\begin{figure}
\figurenum{8}
\plotone{f8.eps}
\caption{ }
\end{figure}

\clearpage
\begin{figure}
\figurenum{9}
\plotone{f9.eps}
\caption{ }
\end{figure}

\clearpage
\begin{figure}
\figurenum{10}
\plotone{f10.eps}
\caption{ }
\end{figure}

\clearpage
\begin{figure}
\figurenum{11}
\plotone{f11.eps}
\caption{ }
\end{figure}

\clearpage
\begin{figure}
\figurenum{12}
\plotone{f12.eps}
\caption{ }
\end{figure}

\clearpage
\begin{figure}
\figurenum{13}
\plotone{f13.eps}
\caption{ }
\end{figure}

\clearpage
\begin{figure}
\figurenum{14}
\plotone{f14.eps}
\caption{ }
\end{figure}

\clearpage
\begin{figure}
\figurenum{15}
\plotone{f15.eps}
\caption{ }
\end{figure}

\clearpage
\begin{figure}
\figurenum{16}
\plotone{f16.eps}
\caption{ }
\end{figure}

\clearpage
\begin{figure}
\figurenum{17}
\epsscale{0.9}
\plotone{f17.eps}
\caption{ }
\end{figure}

\clearpage
\begin{figure}
\figurenum{18}
\epsscale{0.9}
\plotone{f18.eps}
\caption{ }
\end{figure}

\end{document}